\documentclass[aps,prb,twocolumn,groupedaddress,floatfix]{revtex4-1}
\usepackage{graphicx}
\usepackage{xcolor}
\usepackage{float}
\usepackage[utf8]{inputenc}
\usepackage{amsmath}
\usepackage{mathtools}
\DeclareMathOperator{\Tr}{Tr}

\begin{document}

\title{The Raman signal from a hindered hydrogen rotor}

\author{Peter I.C. Cooke$^1$, Ioan B. Magdău$^{1,2}$, Miriam Pena-Alvarez$^1$,  Veronika Afonina$^1$, Philip Dalladay-Simpson$^3$, \\
Xiao-Di Liu$^4$, Ross T. Howie$^3$, Eugene Gregoryanz$^{1,3,4}$ and Graeme J. Ackland$^1$}

\affiliation{$^1$CSEC, SUPA, School of Physics and Astronomy, The University of Edinburgh, Edinburgh EH9 3JZ, United Kingdom}
\affiliation{ $^2$Department of Chemistry and Chemical Engineering, Caltech,  Pasadena, California}
\affiliation{ $^3$Center for High Pressure Science and Technology Advanced Research, Shanghai,China}
\affiliation{ $^4$Key Laboratory of Materials Physics, Institute of Solid State Physics,  CAS, Hefei, China}

\email{gjackland@ed.ac.uk}

\begin{abstract}
We present a method for calculation of Raman modes of the quantum solid phase I solid hydrogen and deuterium. We use the mean-field assumption that the quantised excitations are localized on one molecule.  This is done by  explicit solution of the time-dependent Schroedinger equation in an angle-dependent potential, and direct calculation of the polarisation.
We show that in the free-rotor limit, the H$_2$ and D$_2$ frequencies differ by a factor of 2, which evolves toward $\sqrt{2}$ as the modes acquire librational character due to stronger interactions. The ratio overshoots $\sqrt{2}$ if anharmonic terms weaken the harmonic potential. We also use density functional theory and molecular dynamics to calculate  the E$_{2_g}$ optical phonon frequency and the Raman linewidths.   The molecular dynamics shows that the molecules are not free rotors except at very low pressure and high temperature, and become like oscillators as phase II is approached.
 We fit the interaction strengths to experimental frequencies, but good agreement for intensities requires us to also include strong preferred-orientation and stimulated Raman effects between S$_0$(1) and S$_0$(0) contributions. The experimental Raman spectrum for phase II cannot be reproduced, showing that the mean-field assumption is invalid in that case.
\end{abstract}
\maketitle 

\section{Introduction}
The lowest pressure phase of solid hydrogen comprises a hexagonal close packed (hcp) structure of molecules \cite{Hazen1987,loubeyre1996x,Mao1994,Sharma1980,
  Hemley1990b,Hemley1993,Hardy1968}.  X-ray and neutron studies can
detect the mean nuclear position, but the orientational behavior is
more complicated.  Raman spectroscopy at the lowest pressures, 
shows that the molecules adopt free rotor behavior, characterized by
a series of contributions corresponding to energy levels $J(J+1)$ and
selection rule $\Delta J=2$.  As pressure increases the identification of the single rotational levels become
more complicated, as these low frequency bands significantly broaden\cite{Goncharov1996,Mazin1997,Goncharov1998,Pena2019}.

The free rotor and the simple harmonic oscillator are the two
canonical systems considered in Raman Spectroscopy, but it is
impossible to determine the character of the mode directly from an
experimental peak.  For the diatomic rotor the roton energy levels
are given by:
\begin{equation} E(J) = \frac{\hbar^2}{mr^2}J^2 \label{rotonenergy2D} \end{equation} in 2D and in 3D by:
\begin{equation} E(J) = \frac{\hbar^2}{mr^2}J\left(J+1\right) \label{rotonenergy} \end{equation} where $r$ is the bond length, m is the atomic mass and $J$ is an integer quantum number. The Raman selection rule is $\Delta J = 0, \pm 2$,
where zero corresponds to Rayleigh scattering, $+2$ to Stokes, and -2 to Anti-Stokes processes. This expression holds for both two dimensions (2D) and three dimensions (3D)
rotors, and the energies are fully determined by the bondlength $r$.

For the harmonic oscillator the phonon levels are:
\begin{equation} E(n) = \hbar\omega\left(n+\frac{1}{2}\right)=\sqrt{ \frac{\hbar^2k}{m}} 
\left(n+\frac{1}{2}\right)  \end{equation} 
with $\omega$ the frequency and $k$ the effective spring constant, selection
rules being $\Delta n = 0, \pm 1$. 

A peculiarity of these expressions is the different dependence of energy on
mass.  This becomes particularly relevant when considering the
isotopes of hydrogen, H$_2$ and D$_2$. If one assumes that their electronic
structures are the same, and the Born-Oppenheimer approximation holds,
then at the same density the roton frequencies differ
by a factor of 2, while phonon/libron frequencies differ only by
$\sqrt{2}$. Thus the character of a mode can be determined by
comparing the Raman spectrum of the isotopes.  Experimental studies of
this ratio are presented in the accompanying paper\cite{Pena2019}.

In this paper, we develop the theory for the Raman signal from an
inhibited quantum rotor, assuming that the interactions can be
represented by an external potential.  We illustrate the
principles with a 2D example, then apply it to a 3D case where the
potential will be taken to have the form of interacting quadrupoles and a crystal field on an hcp lattice. To calculate Raman phonon frequencies and to estimate natural linewidths
we use ab initio molecular dynamics simulations.

\section{Theory and Methods}
\subsection{Crossover from roton to libron}

To illustrate the principles, we consider a single mode described by
the Hamiltonian for a 2D hindered rotor in an external potential
$\hat{V}= V_0\cos\theta$:

\begin{equation} \frac{\hbar^2}{mr^2}\frac{\partial^2 \Psi(\theta)}{\partial\theta^2} - V_0\cos\theta\Psi(\theta)=E\Psi(\theta) \end{equation}

\begin{figure}
\includegraphics[width=0.9\linewidth]{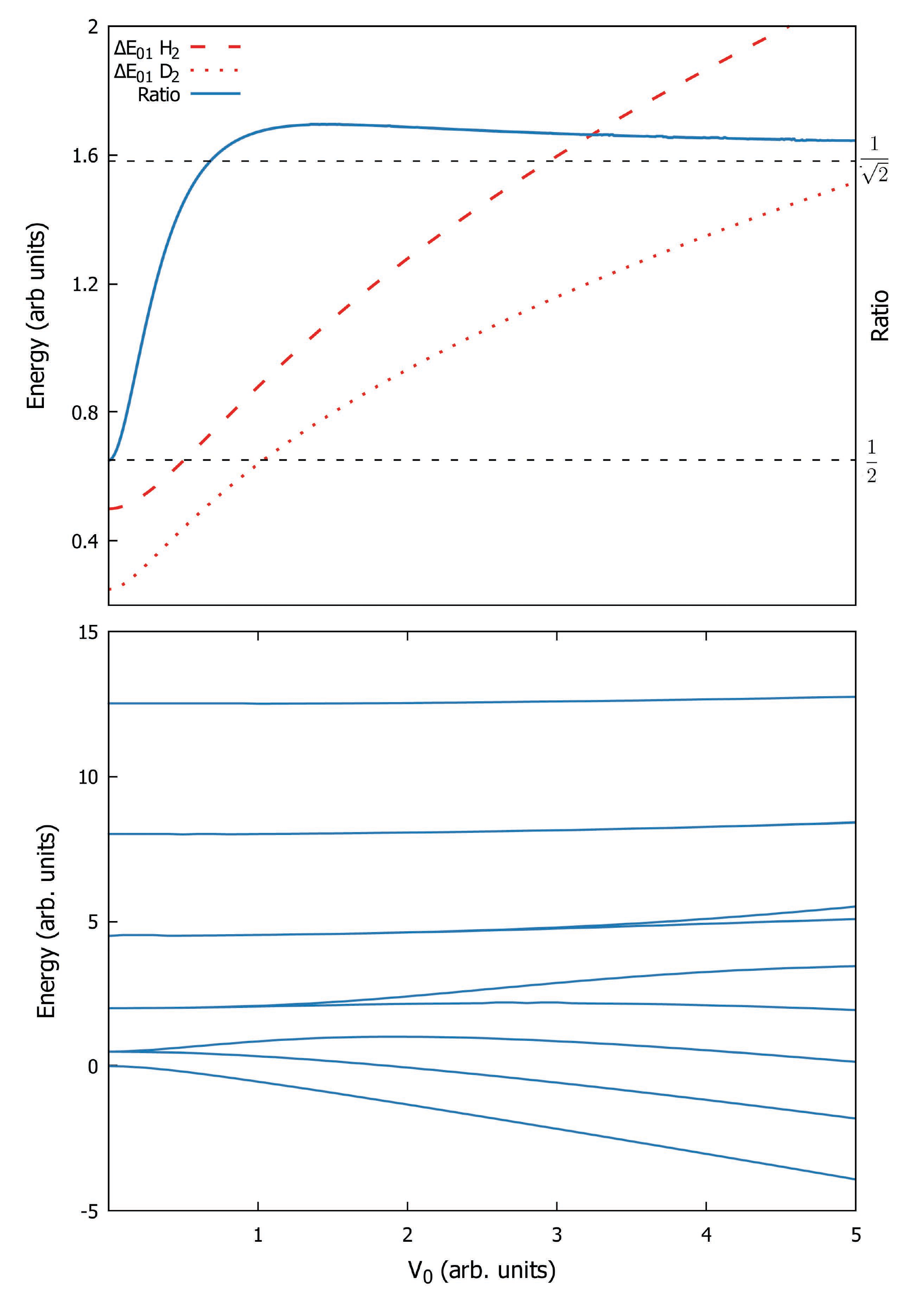}
\caption{2D rotor in a potential of the form $V = V_0cos(\theta)$. Top: left axis shows energy of first
  excitation, compared between $m=1,2$, right axis shows ratio of first excitations. 
Bottom: Energy levels as a function of $V_0$;
energy units are defined by $\frac{\hbar^2}{2m_Hr^2}=1$}
\label{fig:1Dplots}
\end{figure}

The solutions for $E_k(V_0)$ are shown in Figure \ref{fig:1Dplots}.
The frequency ratio for the excitation between rotors with mass 1
(``hydrogen'') and 2 (``deuterium'') is then defined by:
\begin{equation} R =  \frac{E_k(H)-E_i(H)}{E_k(D)-E_i(D)} = \upsilon_{H_2}/\upsilon_{D_2}\end{equation}

where $E_i$ are the calculated energy levels and $\upsilon$ represents
the experimentally-measurable Raman shift.

The limiting cases have $R=2$ for the rotor ($V_0=0$) and $R=\sqrt{2}$
($V_0\rightarrow \infty$), and a surprising result is that $R$ overshoots
and becomes less than its asymptotic value of $\sqrt{2}$: this happens
whenever anharmonic terms make the potential weaker than harmonic at
large distances. Extreme cases for this are the $1/r$ potential 
where the asymptotic value is $R=\frac{1}{2}$ and the purely quartic potential where this ratio becomes $R=2^{2/3}$.

Fig.\ref{fig:1Dplots} also shows that the high
$J$ states remain as free rotors long after the first excited state
passes through the ``oscillator'' value $R=\sqrt{2}$.

\subsection{Spin isomers in solid hydrogen and deuterium}

In solid state hydrogen, the situation becomes more complex.  Below 2 GPa the Raman spectrum can be characterized by a molecular roton spectrum, a lattice
phonon mode, and molecular vibrons at high frequency\cite{VanKranendonk1968}.  At low pressure and temperature the peaks are very sharp, so mode-coupling and perturbative crystal field
splitting is also observable.  \cite{VanKranendonk1983,Pena2019}
Comparing hydrogen and deuterium, the factor $R=2$ 
is observed, demonstrating that the excitations are rotons.

For a free hydrogen molecule the overall wavefunction involves both
nuclear spin and rotational state degrees of freedom.  The nuclear
spin wavefunction can be either a spin-1 symmetric triplet ($ortho$-H)
or spin-0 antisymmetric singlet ($para$-H).  There is no significant
energy associated with nuclear spins, but since protons are fermions with spin $\frac{1}{2}$, the overall wavefunction must be antisymmetric, so only $para$-H can combine with the symmetric $J=0$
rotor ground state.  Consequently, in phase I where intermolecular coupling is weak enough that rotor energy states are localised on a single molecule,
then $para$-H has lower energy than $ortho$-H.  At the phase II boundary, R=$\sqrt{2}$, so the observed excitations are oscillators, not rotors.  $J$ is not a good quantum number, and delocalization of oscillations means that  exchange symmetry does not introduce an independent constraint on each molecule \footnote{The nuclear spin states remain well defined and localised on each molecule}. 
Consequently, $ortho$-H has a higher I-II transition temperature than
$para$-H\cite{lorenzana1989evidence,lorenzana1990orientational}.  
A  broadly similar situation exists in deuterium\cite{silvera1981new}, except that the  deuteron is a spin-1 boson, so $ortho$-D couples
  to the ground state $J=0$, and comprises singlet and quintuplet antisymmetric states as shown in table \ref{tab:symsum}.

These nuclear spin degeneracies result in $ortho-para$ ratios of 3:1 in H$_2$ and 1:2 in $D_2$ at room temperature, which persist metastably on cooling\cite{Strzhemechny2002}.

\begin{table}[H]
\centering
\begin{tabular}{|l|c|c|c|c|c|}
\hline
  & H$_2$-$ortho$ & H$_2$-$para$ & D$_2$-$ortho$ & D$_2$-$para$ & HD \\
\hline
spin symm. & even & odd & even & odd & none \\
spin degen. & 3 & 1 & 6 & 3 & 6\\
rotor symm. & odd & even & even & odd & any  \\ 
rotor state & J=1 & J=0 & J=0 & J=1 & J=0  \\ 
rotor degen. & 3 & 1 & 1 & 3 & 1\\
\hline
\end{tabular}
\caption{Exchange symmetries of wavefunctions showing which nuclear spin states can trap the quantum rotor in high energy $J=1$ state, assuming sufficiently weak interactions that $J$ is a good quantum number.}
\label{tab:symsum}
\end{table}

\subsection{The hindered-rotor Hamiltonian and wavefunction}
Under pressure, intermolecular interactions inhibit
the rotors.  In classical molecular dynamics, this manifests as
increasingly chaotic angular motion of the molecules, while the
molecular centre and bondlengths behave like harmonic oscillators.

To understand the hindered rotor, we  model the
system by describing the rotational motion of a molecule in the
potential of its neighbours on an hcp lattice. Specifically, we solve the angular Schroedinger equation:

\begin{equation}
\begin{aligned}
H(\theta,\phi) = &- \frac{\hbar^2}{mr^2} \left [\frac{1}{\sin \theta} \frac{\partial}{\partial \theta} \left ( \sin \theta \frac{\partial}{ \partial \theta} \right ) + \frac{1}{{\sin^2 \theta}} \frac{\partial^2}{\partial \phi^2} \right]
\\&+ V(\theta, \phi) \label{eq:3D}
\end{aligned}
\end{equation}
where $r=\mathbf{|r|}$ is the molecular bond length and $m$ is the mass of the nucleus.

The potential should have the $P63/mmm$ symmetry of the hcp lattice, and its strength will increase with density. We model it as two distinct contributions, describing the electrostatic and steric interactions.
Long ranged electrostatic interactions, of which quadrupole-quadrupole interactions are dominant, are accounted for by a term with a single parameter, $\lambda$,
\begin{equation}
    V_{e}(\theta, \phi) = \lambda\sum_{i}\left[
    \frac{1}
    { ({\bf R}_i-\frac{\bf r}{2})^5} + \frac{1}{({\bf R}_i+\frac{\bf r}{2})^5} 
    \right]
\end{equation}
Where ${\bf R}_{i}$ is the vector from the central molecule and the $i^{th}$ molecule in the unit cell. The values of $R_{i}$ are taken from the experimental equation of state \cite{loubeyre2013hydrogen} and bondlength $\mathbf{|r|}$ was fitted to the experimental spectra at each pressure and temperature, to within $5\%$ of the gas phase value of $0.74$ \AA. This also affects the moment of inertia, $I=mr^2$, in the kinetic energy term in equation \ref{eq:3D}.

At short range, steric interactions due to Pauli repulsion become important, and quadrupole interactions are enhanced by orientational correlations. We include this by fitting $c_{20}(P)$  and $c_{40}(P)$  directly:
\begin{equation}
    V_{s}(\theta, \phi) = c_{20}Y_{20} + c_{40}Y_{40}
\end{equation}

This approach allows the entire potential $V=V_{e}+V_{s}$ to be described with three parameters: $\lambda,c_{20},c_{40}$.  Interestingly, although $c_{20}$ is allowed in P6$_3$/mmm symmetry, it is zero for central interactions on an hcp lattice with ideal $c/a$ ratio.

 We attempted to include quadrupole correlations at a pairwise level, which gives a parameter-free model.  By neglecting frustration, this strongly overestimates the total quadrupole-quadrupole energy but, surprisingly the angular dependence is too weak to explain the experimental splittings (see Fig. S1).

We expand the potential energy surface $V(\theta,\phi)$ in the basis of spherical harmonics $Y_{kq}(\theta,\phi)$ since these are the solutions to the free rotor problem \cite{Rouvelas2013},
\begin{equation}
V(\theta,\phi)  = \sum_{kq} v_{kq} Y_{kq}(\theta,\phi)
\label{potExp}
\end{equation}
by performing the surface integrals:
\begin{equation}{\label{eq:SHExp}}
v_{kq} = \int Y^{*}_{kq}(\theta',\phi') V(\theta',\phi') \sin(\theta')  d\theta' d\phi'
\end{equation}
We can now express the full Hamiltonian in the basis of the free rotor:
\begin{equation}
H_{lml'm'}^{(0)} = \frac{\hbar^2}{2{\mu}r^2}l(l+1)\delta_{ll'} \delta_{mm'} + V_{lml'm'}
\end{equation}
where the first term is the free rotor kinetic energy and the second is the potential energy operator, expressed as:
\begin{equation}
\begin{aligned}
V_{lml'm'} &= \langle lm | V(\theta,\phi) | l'm' \rangle\\ &= \sum_{kq} v_{kq}  \langle lm |  kq \rangle | l'm' \rangle
\end{aligned}
\end{equation}
where we employed equation \ref{potExp} to expand the potential energy surface. The  $\langle lm |  kq \rangle | l'm' \rangle$ are Clebsch-Gordan coefficients.
 The energy levels are found by diagonalizing the Hamiltonian:
\begin{equation}
H_{nn'}^{(0)} = D^{*}_{n,lm} H_{lml'm'}^{(0)} D_{l'm',n'}
\end{equation}
Note that $l$ and $m$ are no longer good quantum numbers and so we introduce a new quantum number $n$. The new energy levels are $\omega_n = H^{0}_{nn}$, and $D_{l'm',n'}$ is the transformation from the free rotor basis $\lvert lm \rangle$ to the hindered rotor basis $\lvert n \rangle$. The rotational eigenfunctions of the hindered rotor can be evaluated as:
\begin{equation}
\psi_{n} = D^{*}_{n,lm} Y_{lm}
\end{equation}
and their parity (i.e. rotor symmetry) from:
\begin{equation}
\psi_{n}(\theta,\phi) = (-1)^{par}\psi_{n}(\pi-\theta,\phi+\pi)
\end{equation}
Based on the parity we can split the diagonal Hamiltonian into $ortho$ and $para$ contributions:
\begin{equation}
\widehat{H}^{(0)} = \widehat{H}^{(0)}_{o}+\widehat{H}^{(0)}_{p}
\end{equation}
and write the total equilibrium density matrix as:
\begin{equation}
\hat{\rho}^{(0)}=\frac{g_{o}e^{-\widehat{H}^{(0)}_o/kT}+g_{p}e^{-\widehat{H}^{(0)}_p/kT}}{Z(T)}
\end{equation}
where $g_{0}$ and $g_{p}$ are the nuclear spin degeneracies as laid out in table \ref{tab:broadening} and $Z(T) = Z_{o}(T)+Z_{p}(T)$ is the total partition function with the components:
\begin{equation}
\begin{aligned}
Z_o(T) = g_{o}\Tr\left(e^{-\widehat{H}^{(0)}_o/kT}\right) \\
Z_{p}(T) = g_{p}\Tr\left(e^{-\widehat{H}^{(0)}_p/kT}\right)
\end{aligned}
\end{equation}
This assumes equilibration of the $ortho$/$para$ concentrations, however, the nuclear spins equilibrate of the timescale of a typical experiment\cite{Eggert1999}, and so $ortho$-peaks are initially visible even at 10 K. We account for this by redefining the density matrix as:
\begin{equation}
\hat{\rho}^{(0)}=\frac{Z_{o}(T')}{Z(T')}\frac{g_{o}e^{-\widehat{H}^{(0)}_o/kT}}{Z_{o}(T)}+\frac{Z_{p}(T')}{Z(T')}\frac{g_{p}e^{- \widehat{H}^{(0)}_{p}/kT}}{Z_{p}(T)}
\label{eq:$ortho$$para$}
\end{equation}
where we introduced a separate parameter $T'$ that describes the $ortho$-$para$ ratio observed in the experiment as the thermodynamic temperature of the spins;  $T'$ eventually equilibrates to $T$ at a rate which  depends on experimental details.

Now we turn our attention to the polarizability tensor $\widehat{\Pi}_{ij}$. The laser interacts with the system Hamiltonian via a second order field perturbation:
\begin{equation}
\label{eq:FieldPertub}
\widehat{H}(t) = \widehat{H}^{(0)} - E_i(t)\widehat{\Pi}_{ij} E_j(t)
\end{equation}

In the free rotor basis of spherical harmonics, the polarizability tensor can be expressed as:
\begin{equation}
\Pi_{ij,lml'm'} = \langle lm \rvert \mathbf{R}^T(\theta,\phi)\cdot \boldsymbol{\alpha}\cdot\mathbf{R}(\theta,\phi) \lvert l'm' \rangle
\end{equation}
The polarizability tensor depends on the nature of the molecules in the sample. Specifically, for a linear molecule
\begin{equation}
\boldsymbol{\alpha} = 
\left(
\begin{matrix}
1 & 0 & 0 \\
0 & 1 & 0 \\
0 & 0 & \alpha \\
\end{matrix}
\right)
\end{equation}
is the polarizability in the reference frame of the hydrogen molecule. $\alpha$ is a known parameter, taken to have a value of 1.43 from previous experimental work \cite{Bridge1966,Miliordos2018} and considered to be pressure and temperature independent here. The rotation matrix $\mathbf{R}$ transforms the $E$-fields into the frame of the molecule, before they interact with the polarizability ellipsoid. These rotations are effectively averaged in the frame of the single rotor by the orientation probabilities dictated by the wavefunctions. 

Alternatively,  we can express the polarizability tensor in the hindered rotor basis $\lvert n \rangle$:
\begin{equation}
\Pi_{ij,nn'} = D^{*}_{n,lm} \Pi_{ij,lml'm'} D_{l'm',n'}
\end{equation}
by applying the same transformation that diagonalizes the Hamiltonian.

Depending on the orientations of the fields $E_i$ and $E_j$ and the geometry of the experiment, different elements of the tensor will contribute. Raman spectra from diamond anvil cell  experiments are obtained in back-scatter geometry, while the sample normally has preferred orientation along the beam direction. These conditions impose restrictions over which of the $i$ and $j$ components of the polarizability tensor, contribute to the response. The cases we considered are summarized in table \ref{polset}.

\begin{table}[H]
\centering
\begin{tabular}{|c|c|}
\hline
crystal orientation & total response \\
\hline
$c||Z$ & $\sum_{J=X,Y} R\left(\widehat{\Pi}_{XJ}\right)$ \\
isotropic & $\sum_{I,J=X,Y,Z} R\left(\widehat{\Pi}_{IJ}\right)$ \\
\hline
\end{tabular}
\caption{Components of the polarizability that contribute to the total response for each of the crystal orientation.}
\label{polset}
\end{table}

Additionally we suppress all $ortho$ to $para$ transitions by setting the corresponding elements in the transition matrix to zero. 
We only allow transitions that leave the symmetry of the nuclear spin wave-function unchanged.

So far we derived the system Hamiltonian $H^{(0)}_{nn'}$ and the effective polarisability tensor ${\Pi}_{ij,nn'}$ based on the $v_{kq}$ coefficients. 
We have, thus, obtained the energy levels of the hindered rotor, and the transition probabilities between them.\\

\subsection{Calculation of Raman signal}\label{sec:Raman}
We proceed to calculating the actual Raman signal from the response of the quantum system to a sudden excitation. We rely 
on the time-frequency duality to compute the Raman response in the time domain and then obtain the Raman spectrum by Fourier 
transform (FT) of the time response. We achieve this by first propagating the density matrix of the system under the influence of the field and then computing 
the expectation value of the resulting polarization \cite{mukamel1995principles,hamm2011concepts,finneran2016coherent,finneran20172d}. The dynamics is given by the Liouville-von Neumann (LvN) equation:
\begin{equation}
\label{LvN}
\frac{d\hat{\rho}}{dt} = -\frac{i}{\hbar}\left[\widehat{H},\hat{\rho}\right]
\end{equation}
 The advantage of using LvN over the time-dependent Schrodinger equation is that the density matrix can also describe a statistical ensemble of rotors given by:
\begin{equation}
\rho_{nn'} = \sum_{s} p_s \rho^{s}_{nn'}
\end{equation}
where $\rho^{s}_{nn'}$ is the density matrix of the system $s$ and $p_s$ is the probability of finding system $s$. Using the Chain Rule and substituting equation \ref{LvN}, we can express the dynamics of the mixed density matrix $\rho_{nn'}$ as \cite{hamm2011concepts,mukamel1995principles}:
\begin{eqnarray}
\label{RF}
\frac{d\rho_{nn'}}{dt} = \sum_{s} p_s \frac{d\rho^{s}_{nn'}}{dt} +\sum_{s} \frac{dp_s}{dt} \rho^{s}_{nn'} \nonumber \\ = -\frac{i}{\hbar}\left[H,\rho\right]_{nn'} + \sum_{s} \frac{dp_s}{dt} \rho^{s}_{nn'}
\end{eqnarray}
The first term describes the quantum mechanical evolution of the system, while the second term describes the classical statistics and relates to coherence dephasing and energy dissipation. In Redfield formalism, this term can be approximated as:
\begin{equation}
\sum_{s} \frac{dp_s}{dt} \rho^{s}_{nn'}= -\frac{\rho_{nn'}}{\tau}= -\rho_{nn'}\Gamma
\end{equation}
where $\Gamma$ represents the natural line width broadening.
Now we include the total Hamiltonian which contains the external field perturbation, in equation \ref{RF}, and obtain:
\begin{equation}
\label{dyn}
\frac{d\rho_{nn'}}{dt} = - \frac{i}{\hbar}\left[H^{(0)},\rho\right]_{nn'} + \frac{i}{\hbar}\left[\Pi\delta(t),\rho\right]_{nn'} - \rho_{nn'}\Gamma
\end{equation}

Where we assumed the laser field is impulsive and can be treated as a delta function $\delta(t)$. When the field strength is weak and it does not change the original eigenvalues, we can use perturbation theory to describe the evolution of the density matrix \cite{mukamel1995principles,hamm2011concepts}. We write:
\begin{equation}
\rho_{nn'}(t)=\rho^{(0)}_{nn'}+\rho^{(1)}_{nn'}(t)
\end{equation}
where $\hat{\rho}^{(0)}$ is the equilibrium density and $\rho^{(1)}(t)$ describes the response of the system to the external perturbation. Additionally, the equilibrium Hamiltonian is diagonal in the $\lvert n\rangle$ basis, so the first commutator can be easily solved and equation \ref{dyn} becomes:
\begin{equation}
\begin{aligned}
\label{Part}
\frac{d\rho^{(1)}_{nn'}}{dt} = &- \left(i(\omega_n-\omega_{n'})+\Gamma\right) \rho^{(1)}_{nn'}\\ &+ \frac{i}{\hbar}\left[\Pi\delta(t),\rho^{(0)}\right]_{nn'}
\end{aligned}
\end{equation}
We understand this equation intuitively as follows. The equilibrium density matrix $\rho_0$ is diagonal and commutes with the system Hamiltonian and therefore does not contribute to the dynamics. As a result, in the absence of the external perturbation the system is in equilibrium and does not change. The effective polarizability operator acts upon the equilibrium density matrix at $t=0$ and creates off diagonal terms (coherent superpositions of states) $\rho^{(1)}_{nn'}(t=0)$ which then evolve under the system Hamiltonian with oscillating phases $-i(\omega_{n}-\omega_{n'})$ which decay at a rate $\Gamma$. We integrate equation \ref{Part} via a change of variables and obtain:
\begin{equation}
\rho^{(1)}_{nn'}(t) = \frac{i}{\hbar} \int_{-\infty}^{t} \left[\delta(\tau)\Pi,\rho^{(0)}\right]_{nn'} e^{-[i(\omega_n-\omega_{n'})+\Gamma](t-\tau)} d\tau
\end{equation}
and since we assume the perturbation is instantaneous in time, this simplifies to:
\begin{equation}
\rho^{(1)}_{nn'}(t) = \frac{i}{\hbar} \left[\Pi,\rho^{(0)}\right]_{nn'} \label{eq:Widths} e^{-[i(\omega_n-\omega_{n'})+\Gamma]t}
\end{equation}
We discard the second part of the commutator since it is just the complex conjugate of the first part and it carries the same information. The remaining part contains both the Stokes and anti-Stokes Raman contributions. In our energy-sorted basis $\lvert n \rangle$ all Stokes contributions are in the lower triangular matrix and the anti-Stokes are in the upper triangular matrix, so we discard the upper half to keep the pure Stokes signal:
\begin{equation}
\rho^{(1)}_{n,n'<n}(t) = \frac{i}{\hbar} \Pi_{nm}\rho^{(0)}_{mn'} e^{-[i(\omega_n-\omega_{n'})+\Gamma]t}
\end{equation}

Finally, the expectation value of the system polarization expressed in time domain, is:
\begin{equation}
S(t) = {\rm Tr}\left(\Pi_{nm}\rho^{(1)}_{m,n'<m}(t)\right)
\end{equation}
while in frequency domain the observed spectrum is given by:
\begin{equation}
R(f) = \Re{\left( \int_{-\infty}^{\infty}S(t)e^{-i f t} dt \right)}
\label{eq:FFT}
\end{equation}

This gives the entire Raman spectrum with Lorentzian line shapes arising from the broadening $\Gamma$. Pressure broadening gives a similar line shape, so we include this into our simulations by adding a pressure dependent contribution to $\Gamma$:
\begin{equation}
\Gamma = \Gamma^{0}+\Gamma^{P}
\label{eq:Gamma}
\end{equation}
This broadening parameter corresponds to the width of peaks and we calculated it with two approaches. On one hand, the simplest approach is simply to regard this as a fitting parameter, choosing peak widths that match the experimental data. On the other hand, trends in lifetime broadening 
can be calculated from the decay of the angular momentum autocorrelation function extracted from \textit{ab initio} molecular dynamics (AIMD) simulations 
as described in section \ref{sec:AIMD}. There are many approximations in this latter approach, but one robust feature from AIMD is that all timescales are $\sqrt{2}$ longer for deuterium than for hydrogen, so other things being equal the deuterium peaks will be sharper than the hydrogen ones.

\section{Results}

\begin{figure}
\includegraphics[width=1.0\linewidth]{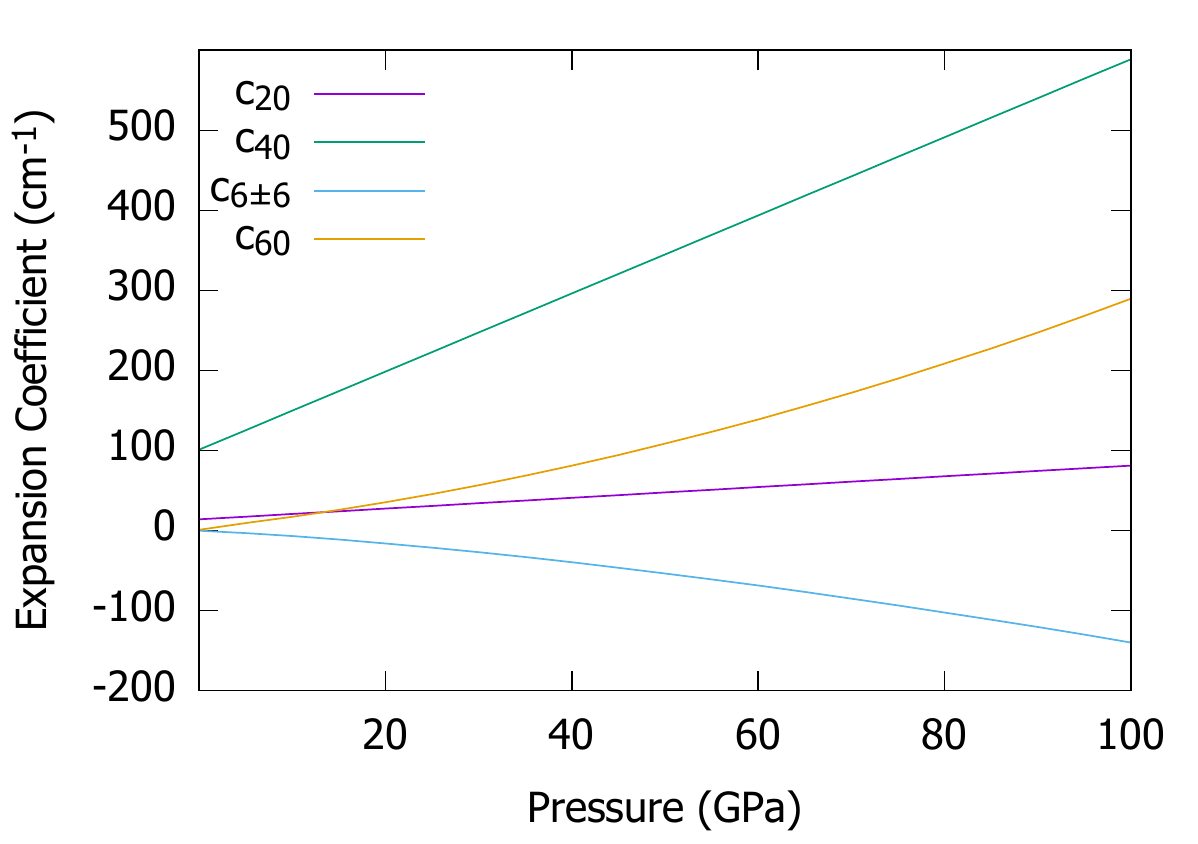}
\caption{Expansion coefficients $v_{lm}$ (Eq. \ref{eq:SHExp}) of spherical harmonics for the potential $V(\theta, \phi)$ (largest 5 are shown with the exception of Y$_{00}$).}
\label{fig:ExpCoeffs}
\end{figure}

\begin{figure}
\includegraphics[width=0.878\linewidth]{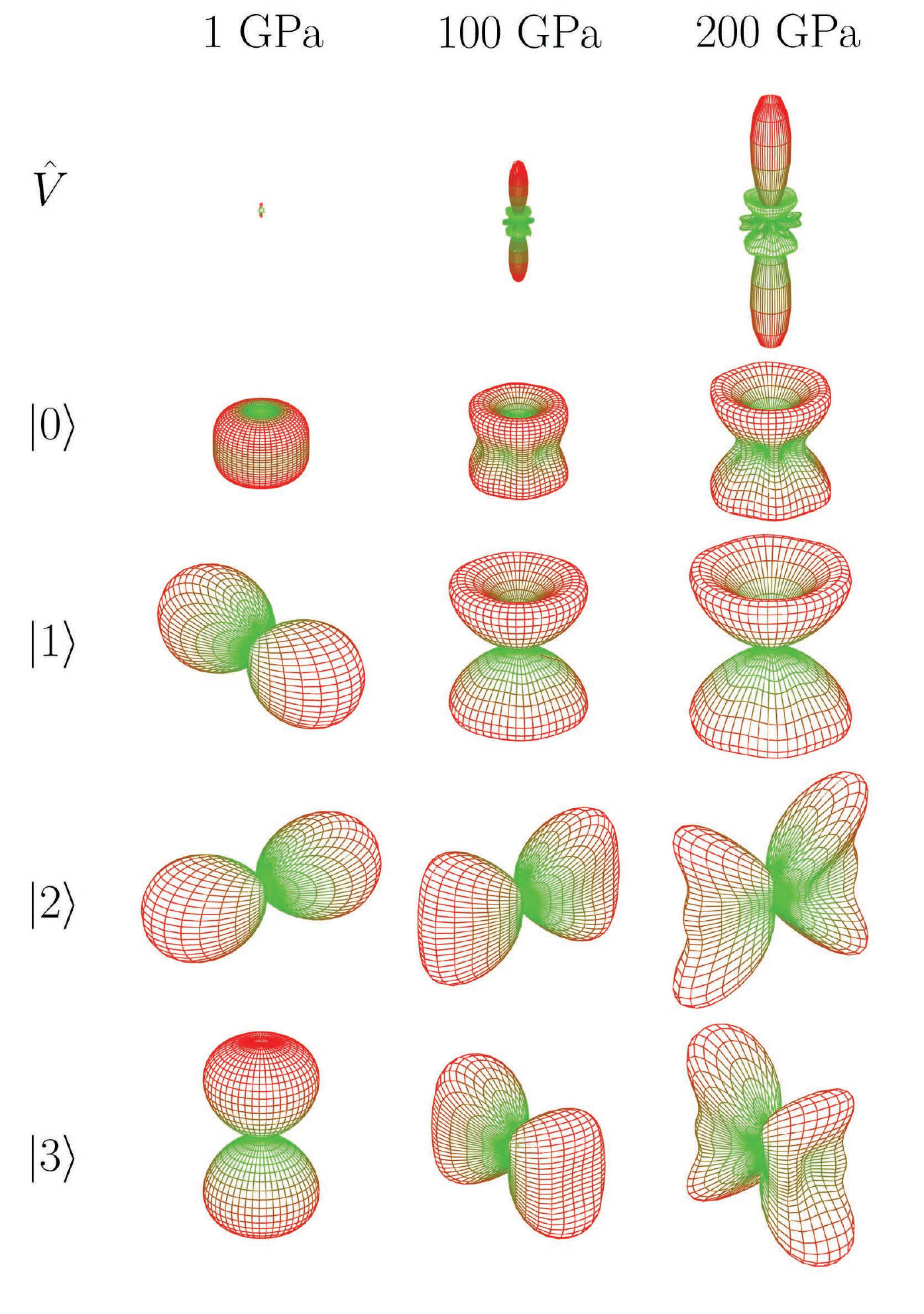}
\caption{3D representation of the potential $V_P(\theta,\phi)$ (top) and four lowest energy wavefunctions $|\Phi(\theta,\phi)|^2$ (below) with increasing 
potential/pressure (left to right) (note the large isotropic $Y_{00}$ component of the potential is not shown here to emphasise the angular dependence).The 
second excited state is doubly degenerate and thus the resulting eigenfunctions from the numerical diagonaliser represent one of many possible 
combinations of basis functions. The bond length was assumed to remain constant at $0.74$ \AA \hspace{1pt} over all pressures as in Fig. \ref{fig:Englevels}.}
\label{fig:wfns}
\end{figure}

\begin{figure}
\includegraphics[width=\linewidth]{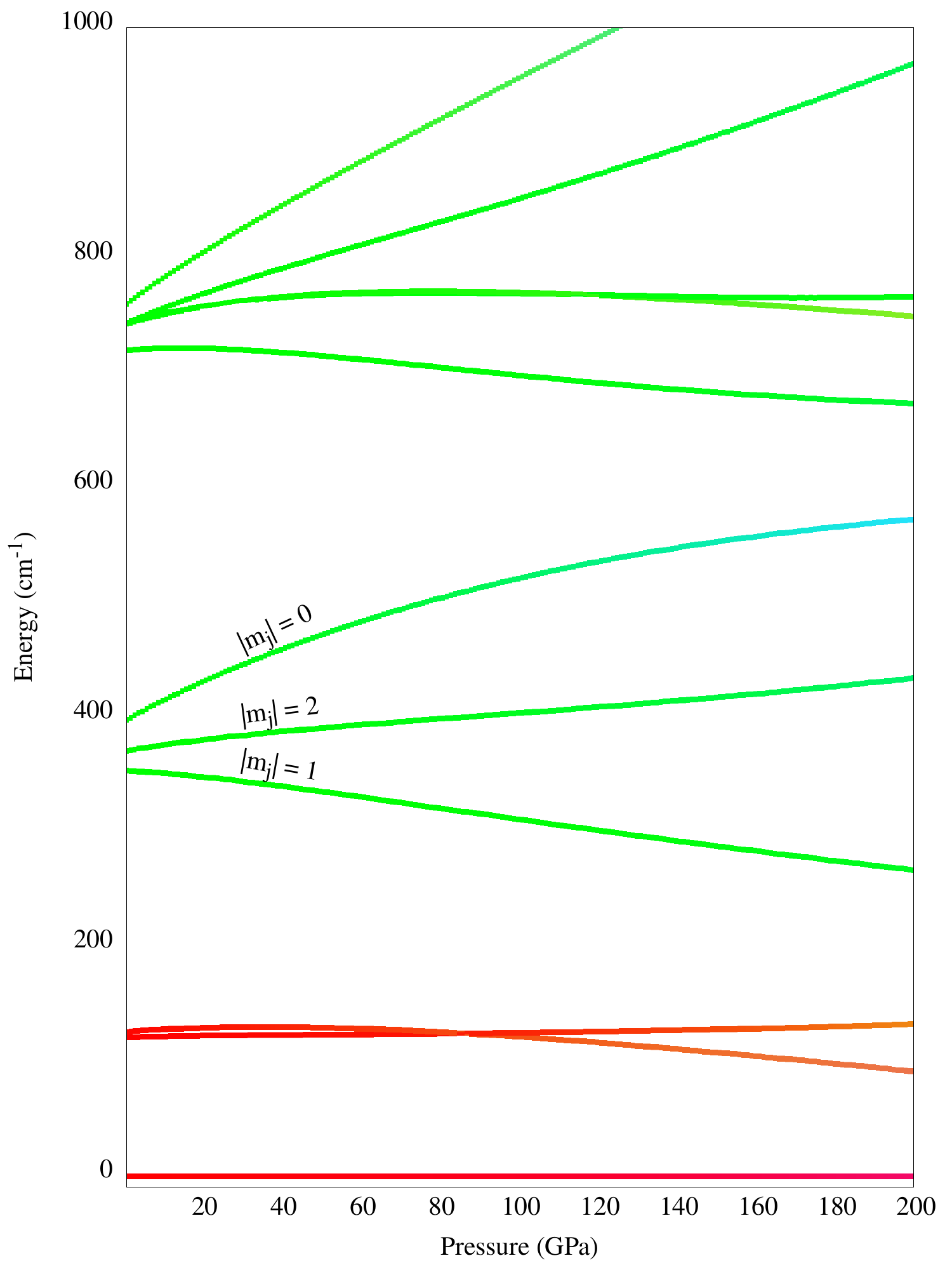}
\caption{Energy levels for a hydrogen rotor in a hexagonal potential: different
  colors indicate the mixing of the spherical harmonics in the
  eigenstate on an RGB scale where red, green and blue pixel values represent contributions from $Y_{00}$ and $Y_{1m}$, $Y_{2m}$ and $Y_{3m}$, $Y_{4m}$ up to $Y_{6m}$ 
  respectively (note: the RGB values have also been rescaled to show $0-10\%$ mixing). For this figure the bond length was assumed to remain constant at $0.74$ \AA \hspace{2pt} as experimental data is not available for the pressures considered here.}
\label{fig:Englevels}
\end{figure}

At low pressure, we obtain ideal rotor behavior, followed by a perturbative region where, e.g. the $J=0\rightarrow 2$ $S_0(0)$ level splits into a triplet.  
The pattern becomes increasingly complicated as pressure increases: not
only are the levels split by the field, but the pure Y$_{lm}$
wavefunctions are mixed, which gives Raman activity to previously
forbidden transitions.  Also, the splitting means a group of low 
energy transitions corresponding to $\Delta J=0$ appear with non-zero shift\cite{silvera1981new}.

For the electrostatic contribution to the potential ($V_{e}$) a single value of $\lambda = 1.98 \times 10^5 \hspace{2pt} $cm$^{-1}$\AA$^{-5}$ was 
used for all pressures and temperatures. Expansion coefficients for the total resultant potential surface $V(\theta,\phi)$  over a range of pressures are 
shown in Fig. \ref{fig:ExpCoeffs}. The same parameters describe both hydrogen and deuterium and are independent of temperature.
Obviously, much better fits can be obtained using more or unphysical  parameters, but doing so could conceal where our single-rotor approximation breaks down. 
This failure is particularly evident in deuterium above $\sim 30$ GPa as phase II emerges (see accompanying paper for details and section \ref{sec:ExpComp} where we show the comparison between our calculated spectra and the experimental ones.). 

Fig. \ref{fig:wfns} shows the potential surface corresponding to the parameters listed above along with the resulting wavefunctions with increasing pressure. At low pressure there is close to zero angular dependence from the potential and the wavefunctions broadly resemble the spherical harmonics. As the pressure is increased up to 100 GPa, minima in the potential surface (shown in green) are seen pointing out of the a-b plane at an angle of $\sim 55^{\circ}$ and at six distinct orientations within the a-b plane. A large maximum in the potential energy surface occurs when the molecule is parallel to the c axis. The emergence of these minima with increasing pressure gives rise to corresponding distortions of the wavefunctions with an increased probability density at $\sim 55^{\circ}$ to the a-b plane seen in the ground and first excited states.  This tendency of the wavefunction to flatten is consistent with AIMD\cite{magdau2013identification,300K}, Monte Carlo\cite{van2020quadrupole} and experiment\cite{ji2019ultrahigh} in phase I, and opposite to theories which predict the molecule pointing preferentially out of plane\cite{freiman1998theory}.
 
Fig. \ref{fig:Englevels} shows the variation of the energy levels with applied potential, with coloring indicating the mixing of spherical harmonics. Relatively little mixing ($5-10\%$) still results in a significant change in the angular dependence of the probability density.

\subsection{Raman mode between split rotational levels}

\begin{figure}
\includegraphics[width=\linewidth]{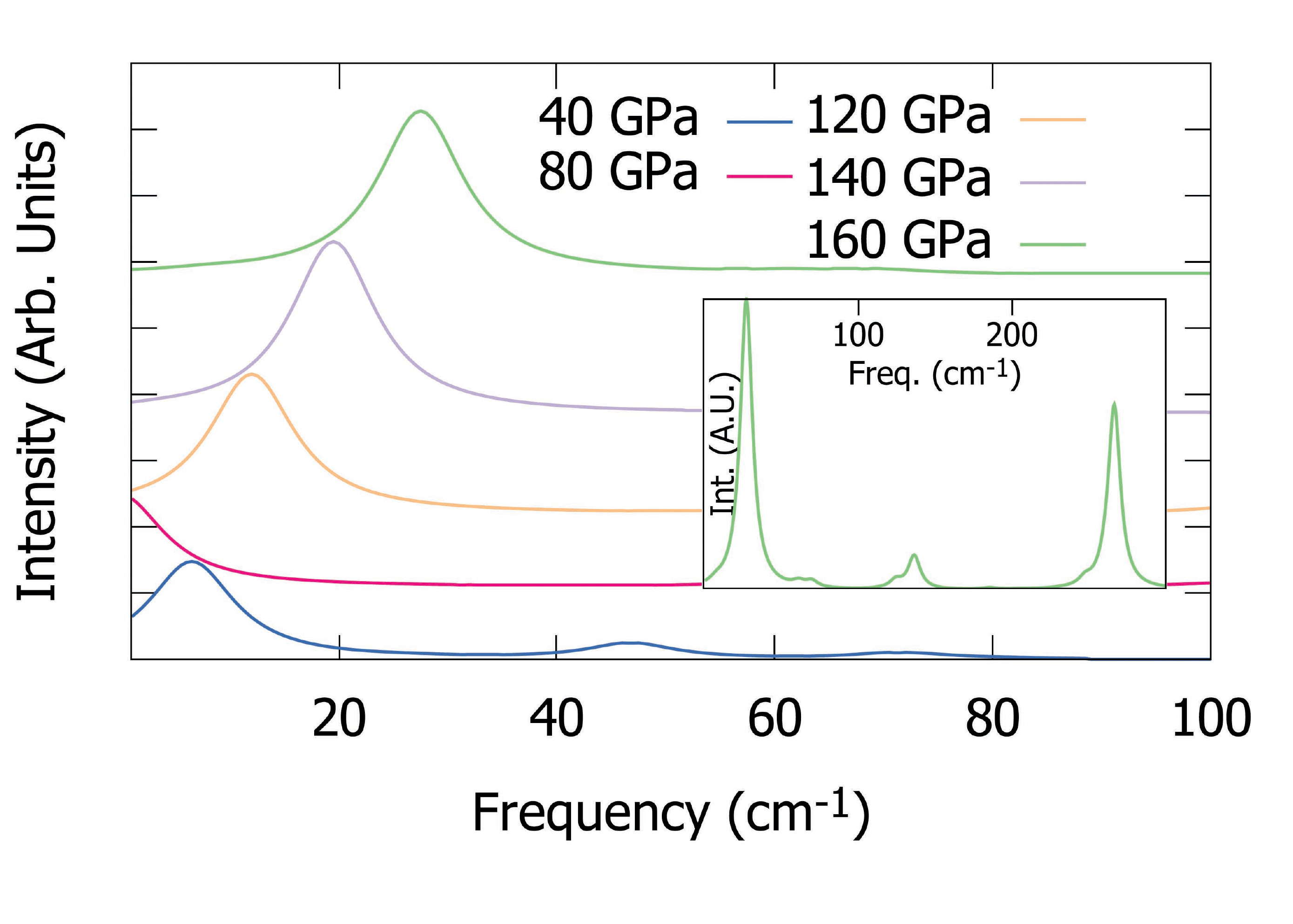}
\caption{Predicted rotational/librational Raman spectrum at frequencies close to the Rayleigh line for hydrogen at 300 K over a range of pressures (broadening parameter is set to $\Gamma = 30$ cm$^{-1}$ for all pressures to allow transitions to be easily identified). Inset shows the $S_0(0)$ triplet and new mode at 160 GPa to demonstrate relative intensities. The bond length was assumed to remain constant at $0.74$ \AA \hspace{2pt} over all pressures. All spectra shown assume a perfect powder measured in backscatter as the reorientational mode is not visible for a sample with c-axis parallel to the incident beam measured in backscatter. At pressures greater than 80 GPa the reorientational mode appears at increasingly larger frequency shifts. At 80 GPa a crossing in the $J=1$ energy levels (seen in Fig. \ref{fig:Englevels}) produces a frequency shift much closer to the Rayleigh line.}
\label{fig:NovelMode}
\end{figure}

Raman modes associated with molecular rotations are typically characterized as librons and rotons. Our calculations show a type of mode which fits neither of these - a reorientational mode.   In the free rotor case, this would be an elastic scattering transition with $\Delta J=0$.   As the potential increases, the Raman shift becomes non-zero: with increasing pressure the low frequency mode  between levels of different $M_J$ emerges from the Rayleigh line (Fig. \ref{fig:NovelMode}).  The selection rule means it can only occur from an initial excited state with $J>0$.  The equivalent mode at zero pressure has been measured using microwave resonance experiments\cite{hardy1975microwave,hardy1977microwave}. At higher pressures the mode may be thought of as the molecule reorienting between inequivalent minima in the potential surface. We note that in the backscattering geometry, with a sample with c axes parallel to the beam, this mode will not be observed.

\section{Molecular dynamics simulations}\label{sec:AIMD}

We have carried out further analysis using a series of ab initio
molecular dynamics (AIMD) simulations in phase I of hydrogen, using methods
presented previously\cite{clark2005first,magdau2013identification,ackland2015appraisal,magdau2013high,magdau2017simple}.

Molecular dynamics of the quantum rotor phase used classical nuclei,
which have long been known to give good results for properties such as
the melting point\cite{bonev2004quantum,liu2012quasi,liu2013anomalous} and to form a basis for a
fully quantum theory.   Zero point
energy favours phase I, but is omitted in AIMD.  Thus the symmetry-breaking phase II of hydrogen is observed even at zero pressure in classical AIMD.  Here we use AIMD to calculate the phonon frequency and to estimate the coherence dephasing parameter which controls our linewidth calculation.

\subsection{Calculation of Linewidths}

 In previous work on vibrational modes, we have shown that
the observed broadening is primarily due to the lifetime of the
mode\cite{magdau2013identification}.  So the parameter $\Gamma$ can be calculated using ab
initio molecular dynamics.  For vibrational modes, the Raman shift can
be extracted directly from the vibrational frequency.  This can be
found by Fourier Transform of the velocity autocorrelation function,
which conveniently also extracts the lifetime broadening from the
anharmonicity.

\begin{figure}
\includegraphics[width=1\columnwidth]{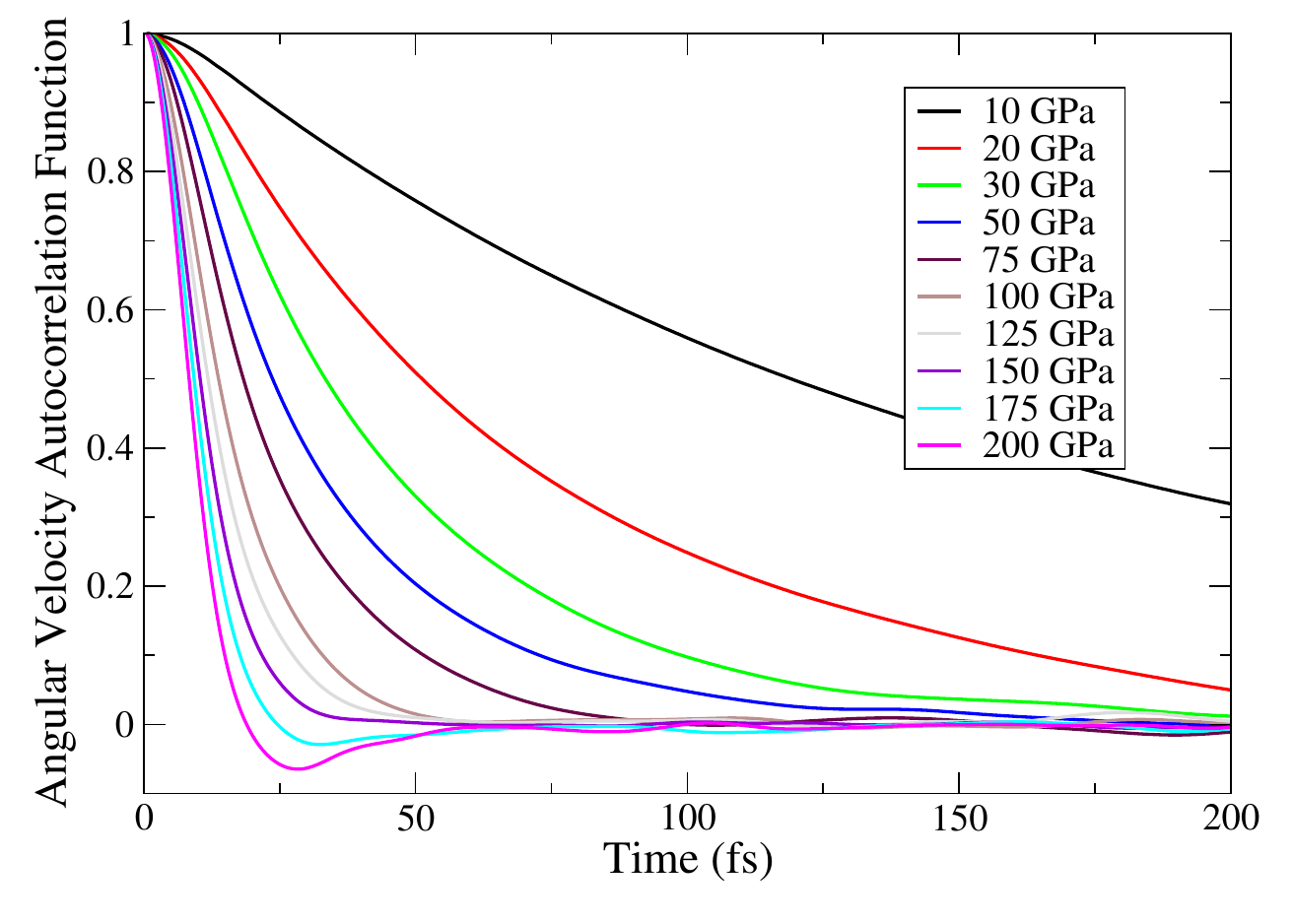}
\caption{Angular momentum autocorrelation functions for diatomic rotors at 300 K and a range of pressures. An ideal rotor has an unchanging correlation function at 1, a harmonic oscillator would have a sinusoidal function between 1 and -1.  Clearly neither is the case here. }
\label{autoP}
\end{figure}

\begin{figure}
\includegraphics[width=1\columnwidth]{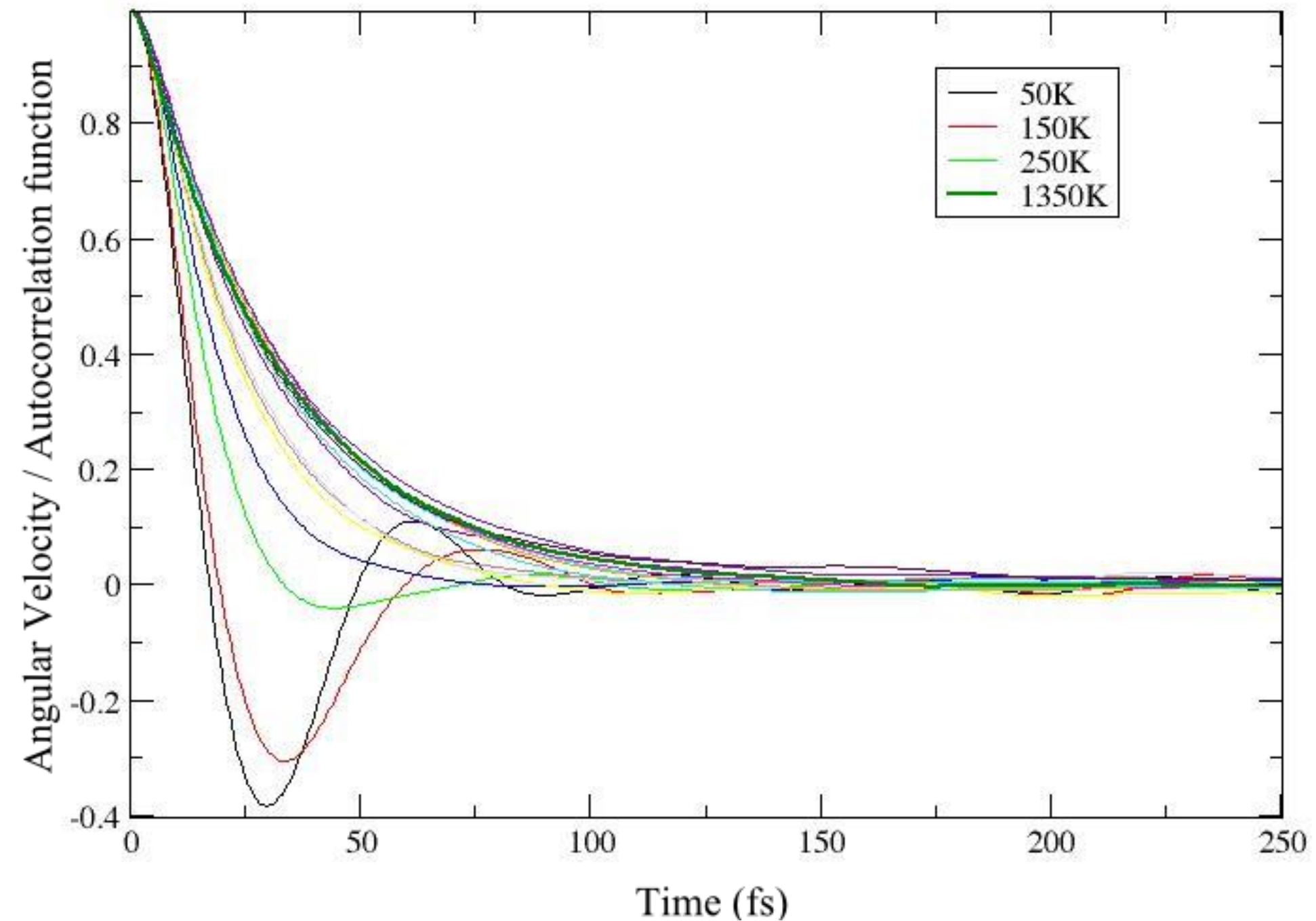}
\caption{Angular momentum correlation functions vs time (fs) at
  100 GPa and a range of temperatures, selected ones labelled, showing convergence on extension into the liquid phase.  The fastest decay is at 250 K,  where all correlation is lost long before a single rotation is  complete.  In phase II (below 150 K, red and black lines), the  autocorrelation becomes negative, indicating that the molecules are librating and not rotating.  }
\label{autoT}
\end{figure}

The simple harmonic oscillator is a special case in that its
quantum energy is directly related to vibrational frequencies (in
Molecular Dynamics) or derivative of the potential energy (in Lattice
Dynamics).  However, the quantised energy levels of a free rotor
(Eq.\ref{rotonenergy}) are unrelated to any classical frequency.  For this reason, the hindered rotor Raman shift cannot be evaluated from AIMD.
However, it is possible to calculate the roton/libron lifetime, and
hence the pressure broadening of the Raman linewidth, from the
autocorrelation function in molecular dynamics (Figs.\ref{autoP} and \ref{autoT}).

From each MD run we identified molecules, and calculated the
autocorrelation function of the angular momentum.

\begin{equation} \ell(t) = \int \frac{\langle \sum_i {\bf L}_i(t-t') {\bf L}_i(t') \rangle}{\langle \sum_i L^2(t') \rangle} dt' \end{equation}

where {\bf L} is the angular momentum, the sum runs over all
molecules and the integral is over the simulation after an 
equilibriation period.

For a rotor, the autocorrelation function decays to zero, while for a
libron there is an anticorrelation period.  In either case, the
classical correlation time is a good proxy for the quantum lifetime,
and the lifetime broadening can be found by Fourier transform of $\ell(t)$. Here the peaks become infinitely sharp in the limit of a perfect rotor or perfect oscillator.

In figure \ref{autoP} we show the autocorrelation as a function of
time for runs at 300 K and pressures up to 200 GPa.  The correlation time
drops to below 100 fs, equivalent to a line broadening of several
hundred cm$^{-1}$. At high pressure, above 175 GPa, we see anticorrelation,
indicative of the short-range freezing-in of the molecular
orientations.  

\begin{figure}
\includegraphics[width=1\columnwidth]{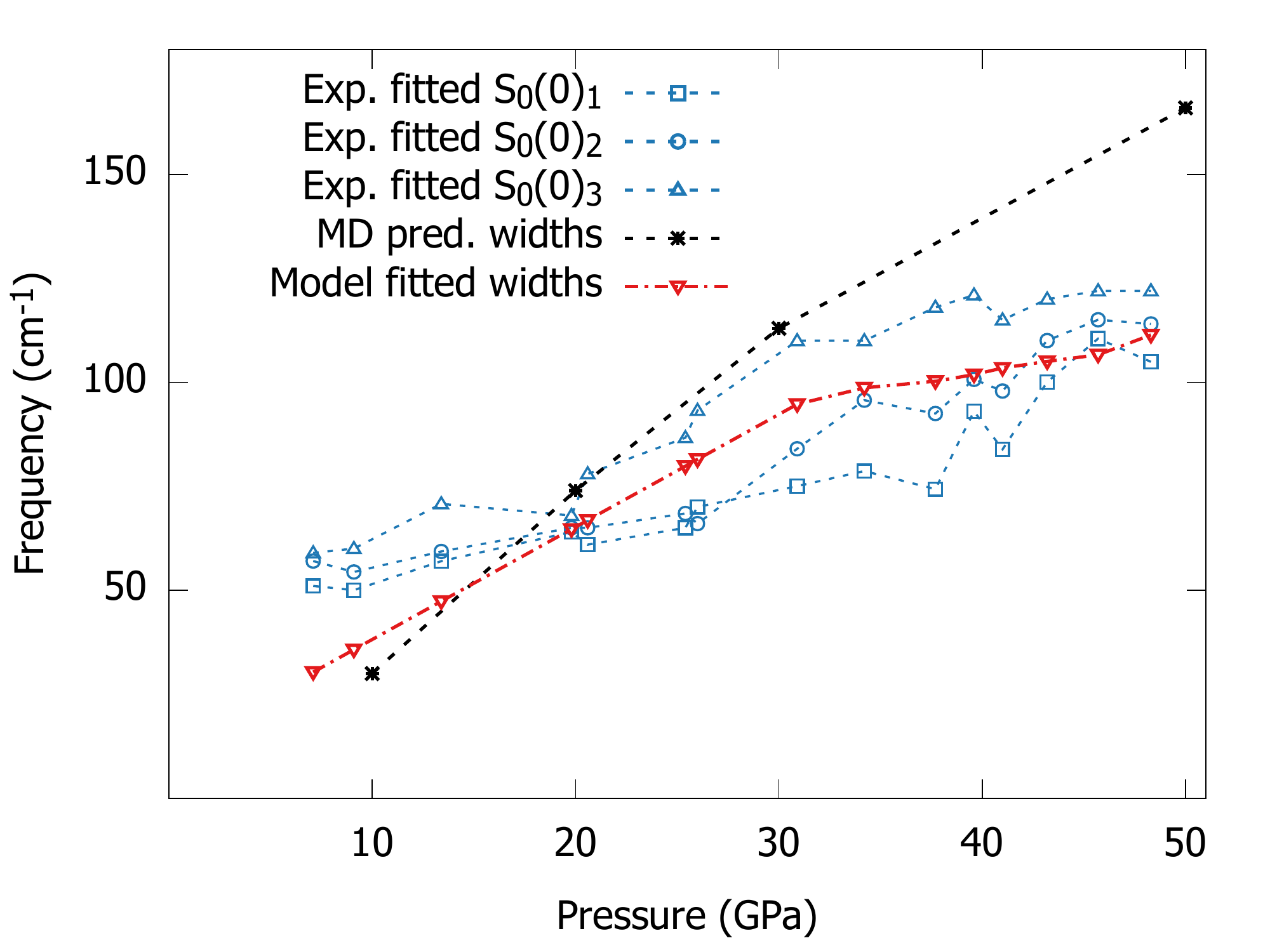}
\caption{Comparison of peak widths for hydrogen at 300 K calculated from statistical fits of Voigt profiles to experimental spectra (blue, note only $S_0(0)$ transitions are shown); MD trajectories with autocorrelation function (black); and single dephasing parameter fitted to entire spectrum in single molecule model (red). The autocorrelation function demonstrates surprisingly good agreement with the fitted values from experiment and the single parameter fitted to the quantum rotor model.}
\label{fig:Widths}
\end{figure}

In Fig. \ref{autoP} we show that the correlation time is highly
reduced with pressure, leading to pressure-broadening of hundreds of
cm$^{-1}$/GPa.  Temperature (Fig. \ref{autoT}) also has an effect, but above 250 K
we find an unusual effect of negative thermal broadening.  Classically, 
this occurs because at high temperature the molecules spin rapidly and the rotation is weakly coupled to other motions. At low temperature, the molecules are strongly coupled, giving a well defined librational harmonic phonon: now the lack of anharmonicity gives the motion a long lifetime. At intermediate temperatures the molecule is neither purely rotating nor librating, so anharmonic coupling leads to rapid decorrelation and consequent reduced lifetime and broadening.
 This is consistent with our experimental observations\cite{Pena2019}, as illustrated in Fig. \ref{fig:Widths}.

\begin{table}[H]
\centering
\begin{tabular}{|c||c|c|c||c|c|c|}
\hline
Pressure & $\tau_{MD}$ & $\tau_{Fit}$ & $\tau_{Exp}$ & $\frac{1}{2\pi}\Gamma_{MD}$  & $\frac{1}{2\pi}\Gamma_{Fit}$ & $\frac{1}{2\pi}\Gamma_{Exp}$ \\
(GPa) & (fs) & (fs) & (fs) & (cm$^{-1}$) & (cm$^{-1}$) & (cm$^{-1}$) \\
\hline
10 &  175 & 149 & 82 & 30 & 36 & 65 \\
20 &  72 & 79 & 65 & 74 & 67 & 82 \\
30 &  47 & 56 & 53 & 113 & 95 & 100 \\
50 &  32 & 48 & 44 & 166 & 103 & 121 \\
\hline
\end{tabular}
\caption{Decorrelation times and associated $\Gamma$  at 300 K derived from AIMD simulation, measured experimental broadening (average of all rotational modes), and best fit of theory to the experimental data.  These data are for hydrogen: assuming the same expectation value for the energy, deuterium decorrelation times will be $\sqrt{2}$ longer.
}
\label{tab:broadening}
\end{table}

\subsection{The phonon mode}

\begin{figure}
\includegraphics[width=\linewidth]{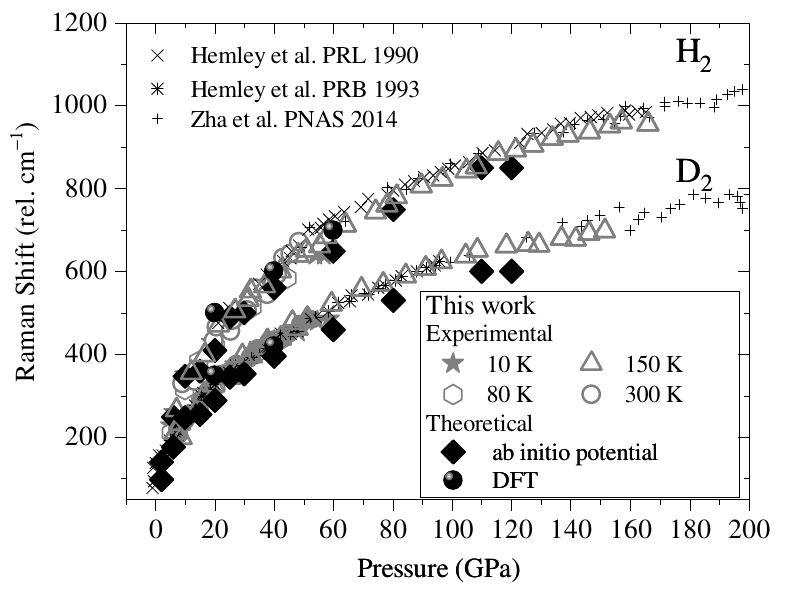}
\caption{Comparison of previous data\cite{Hemley1990,Hemley1993,zha2014raman} the measured\cite{Pena2019} and E$_{2_g}$ calculated phonon modes}
\label{phonon}
\end{figure}

The hcp structure has a single Raman-active E$_{2_g}$ phonon mode.  This
can also be calculated from the MD data by projecting the motion
of the molecular centres onto the wavevector of the Raman-active in
hcp\cite{ackland2014efficacious}.  The phonon has a strong pressure
dependence and extremely good agreement with the experiment can be
seen in Fig. \ref{phonon}.

\section{Comparison with experiment}\label{sec:ExpComp}

We compare our model with the results of high pressure Raman studies.  Details of these experiments are given in the accompanying paper\cite{Pena2019}.

 To compare with experiment, we must further assume that the equation of states are the same for
 hydrogen and deuterium. At relatively low pressures, below 10 GPa
 approximately 5\% difference in specific volume and 10\% in pressure
 has been reported\cite{hemley1990equation}, but later measurements
 suggest the difference is smaller\cite{loubeyre1996x}.
\begin{figure}
\includegraphics[width=1\columnwidth]{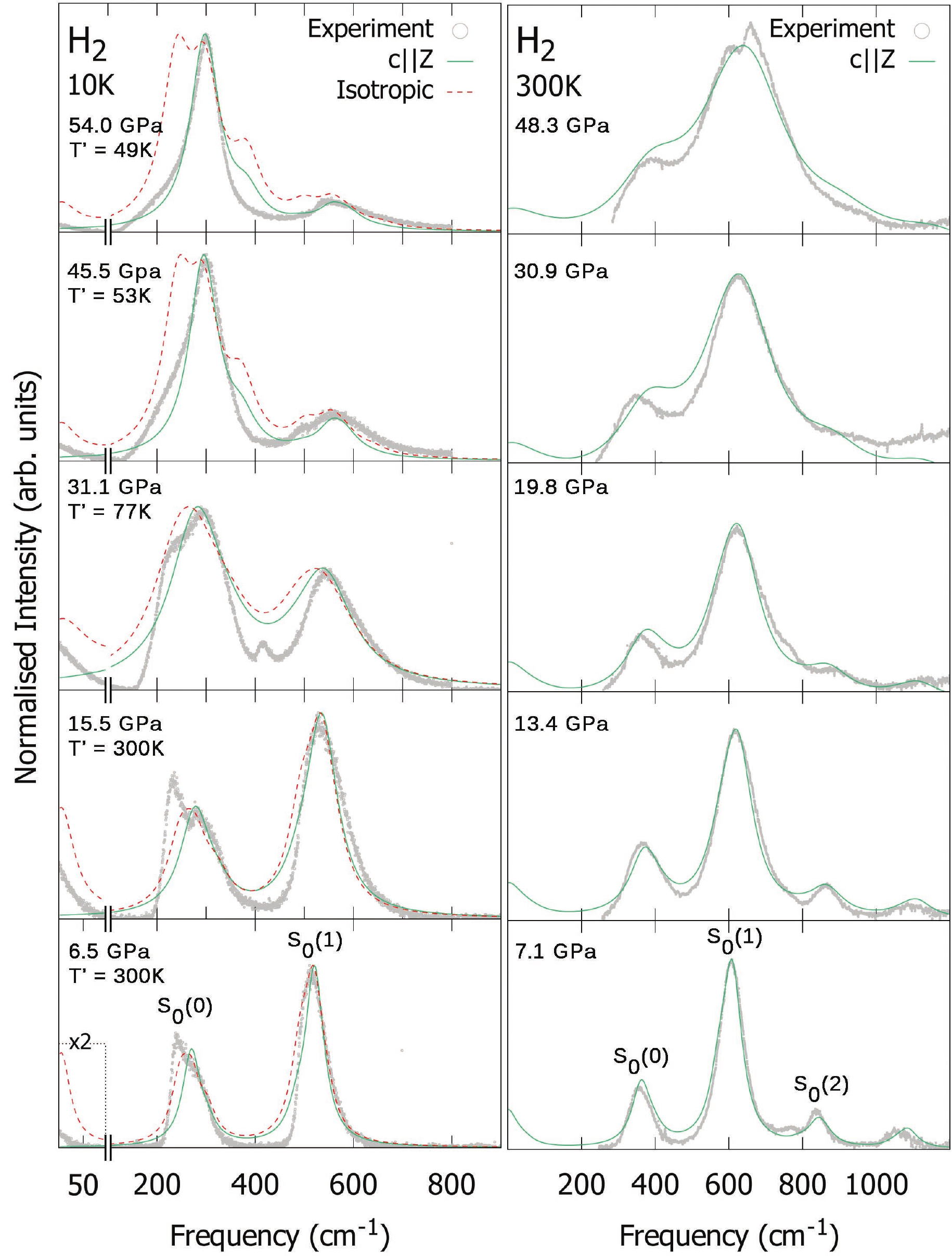}
\caption{Theoretical Raman patterns for H$_2$ at selected
  pressures and temperatures, compared with our experimental data\cite{Pena2019}.
  The pattern comprises peak positions and line-widths calculated from Eq. \ref{eq:FFT} with $\Gamma$ treated as a free parameter (see Fig. \ref{fig:Widths} and table \ref{tab:broadening} for values). Green solid line shows the predicted spectra for a sample with c axis aligned parallel to the beam, the red dashed line shows the predicted spectra in the case of a perfect powder of crystallites. $T'$ denotes the equivalent temperature fitted to the observed $ortho-para$ ratio (see Eq. \ref{eq:$ortho$$para$}).}
\label{fig:H2Spectra}
\end{figure}

\begin{figure}
\includegraphics[width=1\columnwidth]{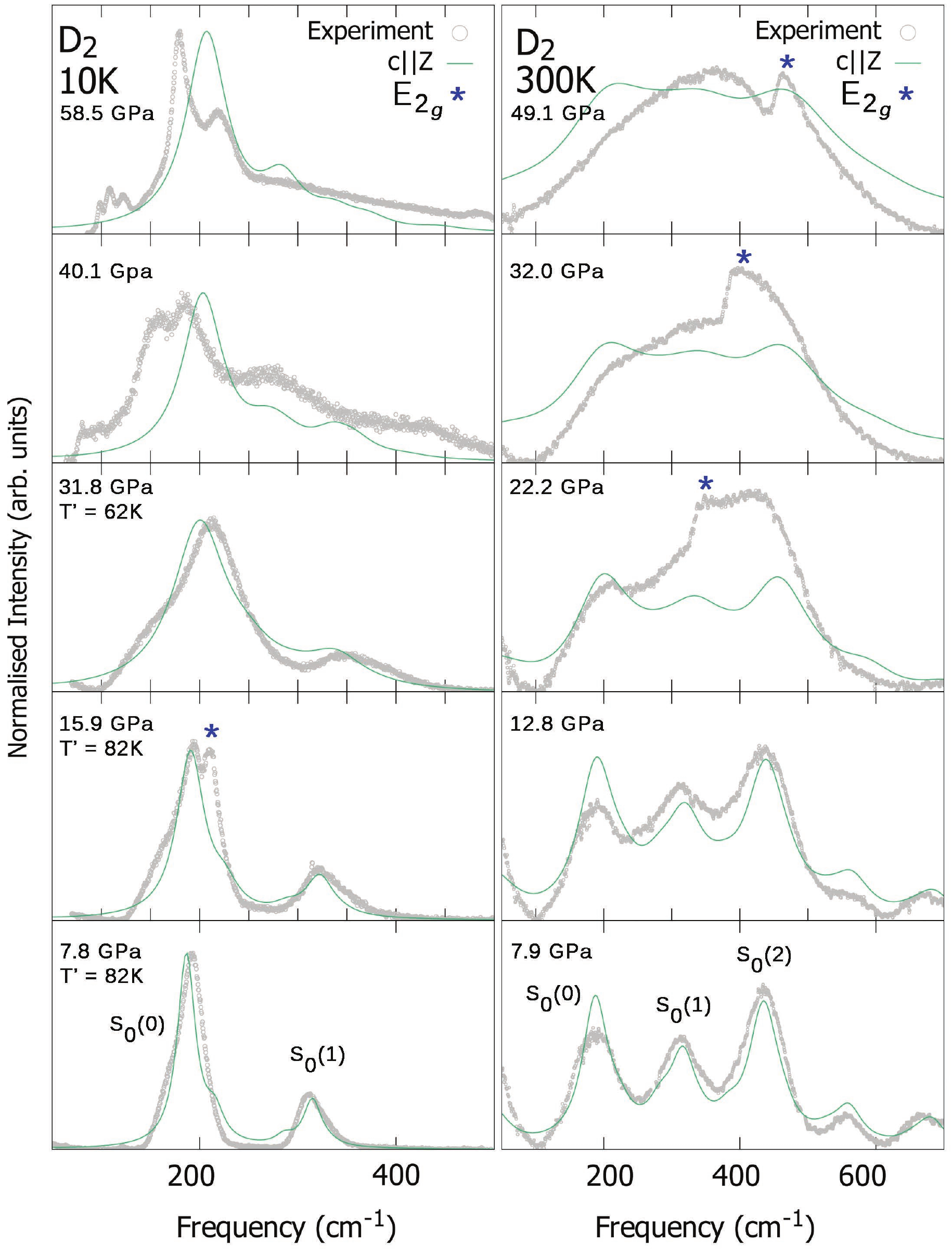}
\caption{Theoretical Raman patterns for D$_2$ at selected
  pressures and temperatures, compared with our experimental data\cite{Pena2019}. The model follows the same procedure for hydrogen described in the figure \ref{fig:H2Spectra} caption with two changes. The mass is increased by a factor of 2 and the parameter $\Gamma$ from Eq. \ref{eq:Widths} is decreased by a factor of $\sqrt{2}$.  The complete mismatch of theory and experiment for 10 K at high pressure indicates the experimental data is for phase II.  the E$_{2_g}$ phonon mode is not included in the calculated spectra}
\label{fig:D2Spectra}
\end{figure}

 The $S_0(0)$ roton peak splits into three ($|\Delta M_J|=0,1,2)$ but this can only be reconciled with the data by  
 noting that in a DAC experiment the crystallites have strong preferred orientation. A back-scattering geometry with the $c$-axis parallel to the beam renders the $\Delta M_J=1$ mode invisible (Table \ref{polset}). This effect is countered by resonant scattering in which the missing $para-$ peaks are enhanced by the absorption and re-emission by the $ortho-$ modes.  Previous work by Eggert et al.\cite{Eggert1999} shows that as $ortho$-H transforms over time, the shape of the $para$-roton peak changes, with the low frequency $S_0(0)_1$ peak eventually disappearing. This non-equilibrium $ortho$-$para$ ratio is described by $T'$, which drops monotonically with time and pressure increase (see Supplemental Materials).

For comparison with experiment for hydrogen at 10 K (figure \ref{fig:H2Spectra}, left panel) two experimental geometries are considered. The green solid line shows the predicted spectra for a sample with c axis parallel to the beam. The dashed red line shows the intensities for a perfect powder. Inspection of the sample suggested that the crystallites are always oriented with the c-axis parallel to the beam as previously seen in X-ray work\cite{akahama2010evidence}. However, whenever significant amounts of $ortho$-H are present, resonant stimulated emission means all three $S_0(0)$ modes have significant intensity\cite{bloembergen1967controlled,VanKranendonk1983,Eggert1999}.  

For deuterium, experimental agreement is good at low pressures ($>$20 GPa). At higher pressures this agreement deteriorates for a number of reasons. The most glaring disagreements seen above 40 GPa at 10 K are caused by the transition to phase II. At 300 K, apparent disagreement is due to the E$_{2_g}$ phonon mode which  appears at similar frequencies to the $S_0(1)$ and $S_2(0)$ peaks (see blue asterisk on figure \ref{fig:D2Spectra}): the phonon is not included in the roton model. We notice a shift upwards in frequency of all modes at higher pressures which could be attributed to a shorter bond length, around 95\% of the gas phase value. 

\section{Discussion and Conclusions}

We have calculated the energy levels and Raman spectra of a perturbed quantum rotor in 2 and 3 dimensions and compared directly with Raman data for high pressure hydrogen.  The 2D data illustrates the isotope effect, with the ratio $\nu_H/\nu_D$ going from 2 to $\sqrt{2}$ as the perturbation becomes stronger, transforming the rotor to a harmonic oscillator. For an anharmonic oscillator, the ratio can be even lower.  In 3D this is more complicated, as there are multiple degenerate minima in the potentials giving different harmonic frequencies.

The Raman spectra are calculated using two distinct approximations: in the traditional approach (see Supplememntal Materials Sec. I), transitions are identified, their Raman intensity calculated and a peakwidth is assigned to each mode. Our alternate approach sets up an excited mixed quantum state, equivalent to linear response textbook Raman theory. We then calculate the polarisation as this mixed state decays according to a {\it single} decorrelation time: Fourier transforming this yields the entire Raman spectrum. Spectra from the two approaches agree very closely (Figs. S3-S6).

A surprisingly good estimate for this decorrelation time can be extracted from the angular momentum autocorrelation function calculated using {\it ab initio} molecular dynamics with classical nuclei.  Using AIMD data for $\Gamma$ and $r$ could eliminate those fitting parameters.

A good angular momentum quantum number, $J$ implies conservation of molecular angular momentum.  The autocorrelation function provides a classical analogy for the concept via the decorrelation time. A good quantum number has infinite decorrelation time and decreasing decorrelation time gives a measure of the "goodness" of the quantum number.  Above 20 GPa, the decorrelation times shown in Fig. \ref{autoP} are less than required for a single, full rotation, and even at low-T and 100 GPa (Fig. \ref{autoT}) scarcely one librational period. Thus the quantum states are well-localised, but are neither good rotors  nor harmonic oscillators.

High pressure hydrogen has a Raman-active phonon mode involving movement of entire layers.  This can be accurately calculated from the AIMD using the projection method (Fig. \ref{phonon}). It is shown to be decoupled from the rotations.

The direct comparison with the entire experimental signal revealed several issues.  Most strikingly, the mean-field theory cannot be made to fit the phase II spectrum, which means that the localised-mode assumptions of the model have broken down: a conclusion also obvious from the molecular dynamics.

In summary, we have calculated the Raman signal from  single-molecule quantum excited states of a perturbed rotor in a hexagonal crystal. We developed a method to directly calculate the entire spectrum with a single decorrelation parameter, which itself can be obtained from ab initio MD calculations. The transformation to the broken-symmetry phase II is clearly signalled by the failure of the theory to explain the data, while a missing peak demonstrates  preferred orientation in the experimental sample.  

The results support the idea that, even within phase I, the motion changes from quantum rotor to quantum libration while the mode remains localised on the molecule.

\begin{acknowledgments} MPA, GJA and EG acknowledge the support of the European Research
Council Grant Hecate Reference No. 695527. GJA acknowledges a Royal Society Wolfson fellowship.
EPSRC funded studentships for PICC, IBM, VA and  computing time (UKCP grant P022561). 
We would like to thank Apurva Dhingra for discussions about this work as part of her Masters thesis.  We thank Gilbert Collins for drawing our attention to the microwave data\cite{hardy1975microwave,hardy1977microwave}. 
\end{acknowledgments}


\begin{thebibliography}{46}
\expandafter\ifx\csname natexlab\endcsname\relax\def\natexlab#1{#1}\fi
\expandafter\ifx\csname bibnamefont\endcsname\relax
  \def\bibnamefont#1{#1}\fi
\expandafter\ifx\csname bibfnamefont\endcsname\relax
  \def\bibfnamefont#1{#1}\fi
\expandafter\ifx\csname citenamefont\endcsname\relax
  \def\citenamefont#1{#1}\fi
\expandafter\ifx\csname url\endcsname\relax
  \def\url#1{\texttt{#1}}\fi
\expandafter\ifx\csname urlprefix\endcsname\relax\def\urlprefix{URL }\fi
\providecommand{\bibinfo}[2]{#2}
\providecommand{\eprint}[2][]{\url{#2}}

\bibitem[{\citenamefont{Hazen et~al.}(1987)\citenamefont{Hazen, Mao, Finger,
  and Hemley}}]{Hazen1987}
\bibinfo{author}{\bibfnamefont{R.~M.} \bibnamefont{Hazen}},
  \bibinfo{author}{\bibfnamefont{H.~K.} \bibnamefont{Mao}},
  \bibinfo{author}{\bibfnamefont{L.~W.} \bibnamefont{Finger}},
  \bibnamefont{and} \bibinfo{author}{\bibfnamefont{R.~J.}
  \bibnamefont{Hemley}}, \bibinfo{journal}{Physical Review B}
  \textbf{\bibinfo{volume}{36}}, \bibinfo{pages}{3944} (\bibinfo{year}{1987}).

\bibitem[{\citenamefont{Loubeyre et~al.}(1996)\citenamefont{Loubeyre,
  LeToullec, Hausermann, Hanfland, Hemley, Mao, and Finger}}]{loubeyre1996x}
\bibinfo{author}{\bibfnamefont{P.}~\bibnamefont{Loubeyre}},
  \bibinfo{author}{\bibfnamefont{R.}~\bibnamefont{LeToullec}},
  \bibinfo{author}{\bibfnamefont{D.}~\bibnamefont{Hausermann}},
  \bibinfo{author}{\bibfnamefont{M.}~\bibnamefont{Hanfland}},
  \bibinfo{author}{\bibfnamefont{R.}~\bibnamefont{Hemley}},
  \bibinfo{author}{\bibfnamefont{H.}~\bibnamefont{Mao}}, \bibnamefont{and}
  \bibinfo{author}{\bibfnamefont{L.}~\bibnamefont{Finger}},
  \bibinfo{journal}{Nature} \textbf{\bibinfo{volume}{383}},
  \bibinfo{pages}{702} (\bibinfo{year}{1996}).

\bibitem[{\citenamefont{Mao and Hemley}(1994)}]{Mao1994}
\bibinfo{author}{\bibfnamefont{H.~K.} \bibnamefont{Mao}} \bibnamefont{and}
  \bibinfo{author}{\bibfnamefont{R.~J.} \bibnamefont{Hemley}},
  \bibinfo{journal}{Reviews of Modern Physics} \textbf{\bibinfo{volume}{66}},
  \bibinfo{pages}{671} (\bibinfo{year}{1994}).

\bibitem[{\citenamefont{Sharma et~al.}(1980)\citenamefont{Sharma, Mao, and
  Bell}}]{Sharma1980}
\bibinfo{author}{\bibfnamefont{K.}~\bibnamefont{Sharma}},
  \bibinfo{author}{\bibfnamefont{H.~K.} \bibnamefont{Mao}}, \bibnamefont{and}
  \bibinfo{author}{\bibfnamefont{P.~M.} \bibnamefont{Bell}},
  \bibinfo{journal}{Physical Review Letters} \textbf{\bibinfo{volume}{44}},
  \bibinfo{pages}{886} (\bibinfo{year}{1980}).

\bibitem[{\citenamefont{Hemley et~al.}(1990{\natexlab{a}})\citenamefont{Hemley,
  Mao, and Shu}}]{Hemley1990b}
\bibinfo{author}{\bibfnamefont{R.~J.} \bibnamefont{Hemley}},
  \bibinfo{author}{\bibfnamefont{H.~K.} \bibnamefont{Mao}}, \bibnamefont{and}
  \bibinfo{author}{\bibfnamefont{J.~F.} \bibnamefont{Shu}},
  \bibinfo{journal}{Physical Review Letters} \textbf{\bibinfo{volume}{65}},
  \bibinfo{pages}{2670} (\bibinfo{year}{1990}{\natexlab{a}}).

\bibitem[{\citenamefont{Hemley et~al.}(1993)\citenamefont{Hemley, Eggert, and
  Mao}}]{Hemley1993}
\bibinfo{author}{\bibfnamefont{R.~J.} \bibnamefont{Hemley}},
  \bibinfo{author}{\bibfnamefont{J.~H.} \bibnamefont{Eggert}},
  \bibnamefont{and} \bibinfo{author}{\bibfnamefont{H.~K.} \bibnamefont{Mao}},
  \bibinfo{journal}{Physical Review B} \textbf{\bibinfo{volume}{48}},
  \bibinfo{pages}{5779} (\bibinfo{year}{1993}).

\bibitem[{\citenamefont{{W. N. Hardy, I. F. Sivlera, K. N.
  Klump}}(1968)}]{Hardy1968}
\bibinfo{author}{\bibfnamefont{O.~S.} \bibnamefont{{W. N. Hardy, I. F. Sivlera,
  K. N. Klump}}}, \textbf{\bibinfo{volume}{21}}, \bibinfo{pages}{291}
  (\bibinfo{year}{1968}).

\bibitem[{\citenamefont{Goncharov et~al.}(1996)\citenamefont{Goncharov, Eggert,
  Mazin, Hemley, and kwang Mao}}]{Goncharov1996}
\bibinfo{author}{\bibfnamefont{A.~F.} \bibnamefont{Goncharov}},
  \bibinfo{author}{\bibfnamefont{J.~H.} \bibnamefont{Eggert}},
  \bibinfo{author}{\bibfnamefont{I.}~\bibnamefont{Mazin}},
  \bibinfo{author}{\bibfnamefont{R.~J.} \bibnamefont{Hemley}},
  \bibnamefont{and} \bibinfo{author}{\bibfnamefont{H.}~\bibnamefont{kwang
  Mao}}, \bibinfo{journal}{Physical Review B - Condensed Matter and Materials
  Physics} \textbf{\bibinfo{volume}{54}}, \bibinfo{pages}{R15590}
  (\bibinfo{year}{1996}).

\bibitem[{\citenamefont{Mazin et~al.}(1997)\citenamefont{Mazin, Hemley,
  Goncharov, Hanfland, and Mao}}]{Mazin1997}
\bibinfo{author}{\bibfnamefont{I.}~\bibnamefont{Mazin}},
  \bibinfo{author}{\bibfnamefont{R.}~\bibnamefont{Hemley}},
  \bibinfo{author}{\bibfnamefont{a.}~\bibnamefont{Goncharov}},
  \bibinfo{author}{\bibfnamefont{M.}~\bibnamefont{Hanfland}}, \bibnamefont{and}
  \bibinfo{author}{\bibfnamefont{H.-k.} \bibnamefont{Mao}},
  \bibinfo{journal}{Physical Review Letters} \textbf{\bibinfo{volume}{78}},
  \bibinfo{pages}{1066} (\bibinfo{year}{1997}).

\bibitem[{\citenamefont{Goncharov et~al.}(1998)\citenamefont{Goncharov, Hemley,
  kwang Mao, and Shu}}]{Goncharov1998}
\bibinfo{author}{\bibfnamefont{A.~F.} \bibnamefont{Goncharov}},
  \bibinfo{author}{\bibfnamefont{R.~J.} \bibnamefont{Hemley}},
  \bibinfo{author}{\bibfnamefont{H.}~\bibnamefont{kwang Mao}},
  \bibnamefont{and} \bibinfo{author}{\bibfnamefont{J.}~\bibnamefont{Shu}},
  \bibinfo{journal}{Physical Review Letters} \textbf{\bibinfo{volume}{80}},
  \bibinfo{pages}{101} (\bibinfo{year}{1998}).

\bibitem[{\citenamefont{Pena-Alvarez et~al.}(2019)\citenamefont{Pena-Alvarez,
  Afonina, Ackland, Dalladay-Simpson, Howie, Liu, and Gregoryanz}}]{Pena2019}
\bibinfo{author}{\bibfnamefont{M.}~\bibnamefont{Pena-Alvarez}},
  \bibinfo{author}{\bibfnamefont{V.}~\bibnamefont{Afonina}},
  \bibinfo{author}{\bibfnamefont{G.~J.} \bibnamefont{Ackland}},
  \bibinfo{author}{\bibfnamefont{P.}~\bibnamefont{Dalladay-Simpson}},
  \bibinfo{author}{\bibfnamefont{R.~T.} \bibnamefont{Howie}},
  \bibinfo{author}{\bibfnamefont{X.-D.} \bibnamefont{Liu}}, \bibnamefont{and}
  \bibinfo{author}{\bibfnamefont{E.}~\bibnamefont{Gregoryanz}},
  \bibinfo{journal}{Physical Review}  (\bibinfo{year}{2019}).

\bibitem[{\citenamefont{{Van Kranendonk} and Karl}(1968)}]{VanKranendonk1968}
\bibinfo{author}{\bibfnamefont{J.}~\bibnamefont{{Van Kranendonk}}}
  \bibnamefont{and} \bibinfo{author}{\bibfnamefont{G.}~\bibnamefont{Karl}},
  \bibinfo{journal}{Reviews of Modern Physics} \textbf{\bibinfo{volume}{40}},
  \bibinfo{pages}{531} (\bibinfo{year}{1968}).

\bibitem[{\citenamefont{{Van Kranendonk}}(1983)}]{VanKranendonk1983}
\bibinfo{author}{\bibfnamefont{J.}~\bibnamefont{{Van Kranendonk}}},
  \emph{\bibinfo{title}{{Solid Hydrogen}}} (\bibinfo{publisher}{Plenun press},
  \bibinfo{address}{new york and london}, \bibinfo{year}{1983}), ISBN
  \bibinfo{isbn}{0-306-41080-X}.

\bibitem[{Note1()}]{Note1}
Note1, \bibinfo{note}{the nuclear spin states remain well defined and localised
  on each molecule}.

\bibitem[{\citenamefont{Lorenzana et~al.}(1989)\citenamefont{Lorenzana,
  Silvera, and Goettel}}]{lorenzana1989evidence}
\bibinfo{author}{\bibfnamefont{H.~E.} \bibnamefont{Lorenzana}},
  \bibinfo{author}{\bibfnamefont{I.~F.} \bibnamefont{Silvera}},
  \bibnamefont{and} \bibinfo{author}{\bibfnamefont{K.~A.}
  \bibnamefont{Goettel}}, \bibinfo{journal}{Phys.Rev.Letters}
  \textbf{\bibinfo{volume}{63}}, \bibinfo{pages}{2080} (\bibinfo{year}{1989}).

\bibitem[{\citenamefont{Lorenzana et~al.}(1990)\citenamefont{Lorenzana,
  Silvera, and Goettel}}]{lorenzana1990orientational}
\bibinfo{author}{\bibfnamefont{H.~E.} \bibnamefont{Lorenzana}},
  \bibinfo{author}{\bibfnamefont{I.~F.} \bibnamefont{Silvera}},
  \bibnamefont{and} \bibinfo{author}{\bibfnamefont{K.~A.}
  \bibnamefont{Goettel}}, \bibinfo{journal}{Physical review letters}
  \textbf{\bibinfo{volume}{64}}, \bibinfo{pages}{1939} (\bibinfo{year}{1990}).

\bibitem[{\citenamefont{Silvera and Wijngaarden}(1981)}]{silvera1981new}
\bibinfo{author}{\bibfnamefont{I.~F.} \bibnamefont{Silvera}} \bibnamefont{and}
  \bibinfo{author}{\bibfnamefont{R.~J.} \bibnamefont{Wijngaarden}},
  \bibinfo{journal}{Phys. Rev. Letters} \textbf{\bibinfo{volume}{47}},
  \bibinfo{pages}{39} (\bibinfo{year}{1981}).

\bibitem[{\citenamefont{Strzhemechny et~al.}(2002)\citenamefont{Strzhemechny,
  Hemley, kwang Mao, Goncharov, and Eggert}}]{Strzhemechny2002}
\bibinfo{author}{\bibfnamefont{M.~A.} \bibnamefont{Strzhemechny}},
  \bibinfo{author}{\bibfnamefont{R.~J.} \bibnamefont{Hemley}},
  \bibinfo{author}{\bibfnamefont{H.}~\bibnamefont{kwang Mao}},
  \bibinfo{author}{\bibfnamefont{A.~F.} \bibnamefont{Goncharov}},
  \bibnamefont{and} \bibinfo{author}{\bibfnamefont{J.~H.}
  \bibnamefont{Eggert}}, \bibinfo{journal}{Physical Review B - Condensed Matter
  and Materials Physics} \textbf{\bibinfo{volume}{66}}, \bibinfo{pages}{1}
  (\bibinfo{year}{2002}).

\bibitem[{\citenamefont{Loubeyre et~al.}(2013)\citenamefont{Loubeyre, Occelli,
  and Dumas}}]{loubeyre2013hydrogen}
\bibinfo{author}{\bibfnamefont{P.}~\bibnamefont{Loubeyre}},
  \bibinfo{author}{\bibfnamefont{F.}~\bibnamefont{Occelli}}, \bibnamefont{and}
  \bibinfo{author}{\bibfnamefont{P.}~\bibnamefont{Dumas}},
  \bibinfo{journal}{Phys. Rev. B} \textbf{\bibinfo{volume}{87}},
  \bibinfo{pages}{134101} (\bibinfo{year}{2013}).

\bibitem[{\citenamefont{Rouvelas}(2013)}]{Rouvelas2013}
\bibinfo{author}{\bibfnamefont{G.~H.} \bibnamefont{Rouvelas}}, Ph.D. thesis,
  \bibinfo{school}{University of Tennessee, Knoxville} (\bibinfo{year}{2013}).

\bibitem[{\citenamefont{Eggert et~al.}(1999)\citenamefont{Eggert, Karmon,
  Hemley, Mao, and Goncharov}}]{Eggert1999}
\bibinfo{author}{\bibfnamefont{J.~H.} \bibnamefont{Eggert}},
  \bibinfo{author}{\bibfnamefont{E.}~\bibnamefont{Karmon}},
  \bibinfo{author}{\bibfnamefont{R.~J.} \bibnamefont{Hemley}},
  \bibinfo{author}{\bibfnamefont{H.-k.} \bibnamefont{Mao}}, \bibnamefont{and}
  \bibinfo{author}{\bibfnamefont{A.~F.} \bibnamefont{Goncharov}},
  \bibinfo{journal}{Proceedings of the National Academy of Sciences of the
  United States of America} \textbf{\bibinfo{volume}{96}},
  \bibinfo{pages}{12269} (\bibinfo{year}{1999}).

\bibitem[{\citenamefont{Bridge and Buckingham}(1966)}]{Bridge1966}
\bibinfo{author}{\bibfnamefont{N.~J.} \bibnamefont{Bridge}} \bibnamefont{and}
  \bibinfo{author}{\bibfnamefont{A.~D.} \bibnamefont{Buckingham}},
  \bibinfo{journal}{Proceedings of the Royal Society of London . Series A}
  \textbf{\bibinfo{volume}{295}}, \bibinfo{pages}{334} (\bibinfo{year}{1966}).

\bibitem[{\citenamefont{Miliordos and Hunt}(2018)}]{Miliordos2018}
\bibinfo{author}{\bibfnamefont{E.}~\bibnamefont{Miliordos}} \bibnamefont{and}
  \bibinfo{author}{\bibfnamefont{K.~L.~C.} \bibnamefont{Hunt}},
  \bibinfo{journal}{The Journal of Chemical Physics}
  \textbf{\bibinfo{volume}{149}}, \bibinfo{pages}{234103}
  (\bibinfo{year}{2018}).

\bibitem[{\citenamefont{Mukamel}(1995)}]{mukamel1995principles}
\bibinfo{author}{\bibfnamefont{S.}~\bibnamefont{Mukamel}},
  \emph{\bibinfo{title}{Principles of nonlinear optical spectroscopy}},
  vol.~\bibinfo{volume}{29} (\bibinfo{publisher}{Oxford university press New
  York}, \bibinfo{year}{1995}).

\bibitem[{\citenamefont{Hamm and Zanni}(2011)}]{hamm2011concepts}
\bibinfo{author}{\bibfnamefont{P.}~\bibnamefont{Hamm}} \bibnamefont{and}
  \bibinfo{author}{\bibfnamefont{M.}~\bibnamefont{Zanni}},
  \emph{\bibinfo{title}{Concepts and methods of 2D infrared spectroscopy}}
  (\bibinfo{publisher}{Cambridge University Press}, \bibinfo{year}{2011}).

\bibitem[{\citenamefont{Finneran et~al.}(2016)\citenamefont{Finneran, Welsch,
  Allodi, Miller, and Blake}}]{finneran2016coherent}
\bibinfo{author}{\bibfnamefont{I.~A.} \bibnamefont{Finneran}},
  \bibinfo{author}{\bibfnamefont{R.}~\bibnamefont{Welsch}},
  \bibinfo{author}{\bibfnamefont{M.~A.} \bibnamefont{Allodi}},
  \bibinfo{author}{\bibfnamefont{T.~F.} \bibnamefont{Miller}},
  \bibnamefont{and} \bibinfo{author}{\bibfnamefont{G.~A.} \bibnamefont{Blake}},
  \bibinfo{journal}{Proceedings of the National Academy of Sciences}
  \textbf{\bibinfo{volume}{113}}, \bibinfo{pages}{6857} (\bibinfo{year}{2016}).

\bibitem[{\citenamefont{Finneran et~al.}(2017)\citenamefont{Finneran, Welsch,
  Allodi, Miller, and Blake}}]{finneran20172d}
\bibinfo{author}{\bibfnamefont{I.~A.} \bibnamefont{Finneran}},
  \bibinfo{author}{\bibfnamefont{R.}~\bibnamefont{Welsch}},
  \bibinfo{author}{\bibfnamefont{M.~A.} \bibnamefont{Allodi}},
  \bibinfo{author}{\bibfnamefont{T.~F.} \bibnamefont{Miller}},
  \bibnamefont{and} \bibinfo{author}{\bibfnamefont{G.~A.} \bibnamefont{Blake}},
  \bibinfo{journal}{Journal of Physical Chemistry Letters}
  \textbf{\bibinfo{volume}{8}}, \bibinfo{pages}{4640} (\bibinfo{year}{2017}).

\bibitem[{\citenamefont{Magd{\u{a}}u and
  Ackland}(2013)}]{magdau2013identification}
\bibinfo{author}{\bibfnamefont{I.~B.} \bibnamefont{Magd{\u{a}}u}}
  \bibnamefont{and} \bibinfo{author}{\bibfnamefont{G.~J.}
  \bibnamefont{Ackland}}, \bibinfo{journal}{Phys. Rev. B}
  \textbf{\bibinfo{volume}{87}}, \bibinfo{pages}{174110}
  (\bibinfo{year}{2013}).

\bibitem[{\citenamefont{Ackland and Loveday}(2019)}]{300K}
\bibinfo{author}{\bibfnamefont{G.}~\bibnamefont{Ackland}} \bibnamefont{and}
  \bibinfo{author}{\bibfnamefont{J.}~\bibnamefont{Loveday}},
  \bibinfo{journal}{Physical Review B, submitted}
  \textbf{\bibinfo{volume}{https://arxiv.org/abs/1910.05260}}
  (\bibinfo{year}{2019}).

\bibitem[{\citenamefont{van~de Bund and Ackland}(2020)}]{van2020quadrupole}
\bibinfo{author}{\bibfnamefont{S.}~\bibnamefont{van~de Bund}} \bibnamefont{and}
  \bibinfo{author}{\bibfnamefont{G.~J.} \bibnamefont{Ackland}},
  \bibinfo{journal}{Physical Review B} \textbf{\bibinfo{volume}{101}},
  \bibinfo{pages}{014103} (\bibinfo{year}{2020}).

\bibitem[{\citenamefont{Ji et~al.}(2019)\citenamefont{Ji, Li, Liu, Smith,
  Majumdar, Luo, Ahuja, Shu, Wang, Sinogeikin et~al.}}]{ji2019ultrahigh}
\bibinfo{author}{\bibfnamefont{C.}~\bibnamefont{Ji}},
  \bibinfo{author}{\bibfnamefont{B.}~\bibnamefont{Li}},
  \bibinfo{author}{\bibfnamefont{W.}~\bibnamefont{Liu}},
  \bibinfo{author}{\bibfnamefont{J.~S.} \bibnamefont{Smith}},
  \bibinfo{author}{\bibfnamefont{A.}~\bibnamefont{Majumdar}},
  \bibinfo{author}{\bibfnamefont{W.}~\bibnamefont{Luo}},
  \bibinfo{author}{\bibfnamefont{R.}~\bibnamefont{Ahuja}},
  \bibinfo{author}{\bibfnamefont{J.}~\bibnamefont{Shu}},
  \bibinfo{author}{\bibfnamefont{J.}~\bibnamefont{Wang}},
  \bibinfo{author}{\bibfnamefont{S.}~\bibnamefont{Sinogeikin}},
  \bibnamefont{et~al.}, \bibinfo{journal}{Nature}
  \textbf{\bibinfo{volume}{573}}, \bibinfo{pages}{558} (\bibinfo{year}{2019}).

\bibitem[{\citenamefont{Freiman et~al.}(1998)\citenamefont{Freiman, Tretyak,
  and Je{\.z}owski}}]{freiman1998theory}
\bibinfo{author}{\bibfnamefont{Y.~A.} \bibnamefont{Freiman}},
  \bibinfo{author}{\bibfnamefont{S.}~\bibnamefont{Tretyak}}, \bibnamefont{and}
  \bibinfo{author}{\bibfnamefont{A.}~\bibnamefont{Je{\.z}owski}},
  \bibinfo{journal}{Journal of low temperature physics}
  \textbf{\bibinfo{volume}{111}}, \bibinfo{pages}{475} (\bibinfo{year}{1998}).

\bibitem[{\citenamefont{Hardy and Berlinsky}(1975)}]{hardy1975microwave}
\bibinfo{author}{\bibfnamefont{W.}~\bibnamefont{Hardy}} \bibnamefont{and}
  \bibinfo{author}{\bibfnamefont{A.}~\bibnamefont{Berlinsky}},
  \bibinfo{journal}{Physical Review Letters} \textbf{\bibinfo{volume}{34}},
  \bibinfo{pages}{1520} (\bibinfo{year}{1975}).

\bibitem[{\citenamefont{Hardy et~al.}(1977)\citenamefont{Hardy, Berlinsky, and
  Harris}}]{hardy1977microwave}
\bibinfo{author}{\bibfnamefont{W.}~\bibnamefont{Hardy}},
  \bibinfo{author}{\bibfnamefont{A.}~\bibnamefont{Berlinsky}},
  \bibnamefont{and} \bibinfo{author}{\bibfnamefont{A.}~\bibnamefont{Harris}},
  \bibinfo{journal}{Canadian Journal of Physics} \textbf{\bibinfo{volume}{55}},
  \bibinfo{pages}{1150} (\bibinfo{year}{1977}).

\bibitem[{\citenamefont{Clark et~al.}(2005)\citenamefont{Clark, Segall,
  Pickard, Hasnip, Probert, Refson, and Payne}}]{clark2005first}
\bibinfo{author}{\bibfnamefont{S.~J.} \bibnamefont{Clark}},
  \bibinfo{author}{\bibfnamefont{M.~D.} \bibnamefont{Segall}},
  \bibinfo{author}{\bibfnamefont{C.~J.} \bibnamefont{Pickard}},
  \bibinfo{author}{\bibfnamefont{P.~J.} \bibnamefont{Hasnip}},
  \bibinfo{author}{\bibfnamefont{M.~I.} \bibnamefont{Probert}},
  \bibinfo{author}{\bibfnamefont{K.}~\bibnamefont{Refson}}, \bibnamefont{and}
  \bibinfo{author}{\bibfnamefont{M.~C.} \bibnamefont{Payne}},
  \bibinfo{journal}{Zeitschrift f{\"u}r Kristallographie-Crystalline Materials}
  \textbf{\bibinfo{volume}{220}}, \bibinfo{pages}{567} (\bibinfo{year}{2005}).

\bibitem[{\citenamefont{Ackland and Magd{\u{a}}u}(2015)}]{ackland2015appraisal}
\bibinfo{author}{\bibfnamefont{G.~J.} \bibnamefont{Ackland}} \bibnamefont{and}
  \bibinfo{author}{\bibfnamefont{I.~B.} \bibnamefont{Magd{\u{a}}u}},
  \bibinfo{journal}{Cogent Physics} \textbf{\bibinfo{volume}{2}},
  \bibinfo{pages}{1049477} (\bibinfo{year}{2015}).

\bibitem[{\citenamefont{Magd{\u{a}}u and Ackland}(2014)}]{magdau2013high}
\bibinfo{author}{\bibfnamefont{I.~B.} \bibnamefont{Magd{\u{a}}u}}
  \bibnamefont{and} \bibinfo{author}{\bibfnamefont{G.~J.}
  \bibnamefont{Ackland}}, \bibinfo{journal}{J. Phys.: Conf. Ser.}
  \textbf{\bibinfo{volume}{500}}, \bibinfo{pages}{032012}
  (\bibinfo{year}{2014}).

\bibitem[{\citenamefont{Magd{\u{a}}u et~al.}(2017)\citenamefont{Magd{\u{a}}u,
  Marques, Borgulya, and Ackland}}]{magdau2017simple}
\bibinfo{author}{\bibfnamefont{I.~B.} \bibnamefont{Magd{\u{a}}u}},
  \bibinfo{author}{\bibfnamefont{M.}~\bibnamefont{Marques}},
  \bibinfo{author}{\bibfnamefont{B.}~\bibnamefont{Borgulya}}, \bibnamefont{and}
  \bibinfo{author}{\bibfnamefont{G.~J.} \bibnamefont{Ackland}},
  \bibinfo{journal}{Phys.Rev.B} \textbf{\bibinfo{volume}{95}},
  \bibinfo{pages}{094107} (\bibinfo{year}{2017}).

\bibitem[{\citenamefont{Bonev et~al.}(2004)\citenamefont{Bonev, Schwegler,
  Ogitsu, and Galli}}]{bonev2004quantum}
\bibinfo{author}{\bibfnamefont{S.~A.} \bibnamefont{Bonev}},
  \bibinfo{author}{\bibfnamefont{E.}~\bibnamefont{Schwegler}},
  \bibinfo{author}{\bibfnamefont{T.}~\bibnamefont{Ogitsu}}, \bibnamefont{and}
  \bibinfo{author}{\bibfnamefont{G.}~\bibnamefont{Galli}},
  \bibinfo{journal}{Nature} \textbf{\bibinfo{volume}{431}},
  \bibinfo{pages}{669} (\bibinfo{year}{2004}).

\bibitem[{\citenamefont{Liu et~al.}(2012)\citenamefont{Liu, Wang, and
  Ma}}]{liu2012quasi}
\bibinfo{author}{\bibfnamefont{H.}~\bibnamefont{Liu}},
  \bibinfo{author}{\bibfnamefont{H.}~\bibnamefont{Wang}}, \bibnamefont{and}
  \bibinfo{author}{\bibfnamefont{Y.}~\bibnamefont{Ma}},
  \bibinfo{journal}{Journal of Physical Chemistry C}
  \textbf{\bibinfo{volume}{116}}, \bibinfo{pages}{9221} (\bibinfo{year}{2012}).

\bibitem[{\citenamefont{Liu et~al.}(2013)\citenamefont{Liu, Hernandez, Yan, and
  Ma}}]{liu2013anomalous}
\bibinfo{author}{\bibfnamefont{H.}~\bibnamefont{Liu}},
  \bibinfo{author}{\bibfnamefont{E.~R.} \bibnamefont{Hernandez}},
  \bibinfo{author}{\bibfnamefont{J.}~\bibnamefont{Yan}}, \bibnamefont{and}
  \bibinfo{author}{\bibfnamefont{Y.}~\bibnamefont{Ma}},
  \bibinfo{journal}{Journal of Physical Chemistry C}
  \textbf{\bibinfo{volume}{117}}, \bibinfo{pages}{11873}
  (\bibinfo{year}{2013}).

\bibitem[{\citenamefont{Ackland and
  Magd{\u{a}}u}(2014)}]{ackland2014efficacious}
\bibinfo{author}{\bibfnamefont{G.~J.} \bibnamefont{Ackland}} \bibnamefont{and}
  \bibinfo{author}{\bibfnamefont{I.~B.} \bibnamefont{Magd{\u{a}}u}},
  \bibinfo{journal}{High Pressure Research} \textbf{\bibinfo{volume}{34}},
  \bibinfo{pages}{198} (\bibinfo{year}{2014}).

\bibitem[{\citenamefont{Hemley et~al.}(1990{\natexlab{b}})\citenamefont{Hemley,
  Mao, and Shu}}]{Hemley1990}
\bibinfo{author}{\bibfnamefont{R.~J.} \bibnamefont{Hemley}},
  \bibinfo{author}{\bibfnamefont{H.~K.} \bibnamefont{Mao}}, \bibnamefont{and}
  \bibinfo{author}{\bibfnamefont{J.~F.} \bibnamefont{Shu}},
  \bibinfo{journal}{Physical Review Letters} \textbf{\bibinfo{volume}{65}},
  \bibinfo{pages}{2670} (\bibinfo{year}{1990}{\natexlab{b}}).

\bibitem[{\citenamefont{Zha et~al.}(2014)\citenamefont{Zha, Cohen, Mao, and
  Hemley}}]{zha2014raman}
\bibinfo{author}{\bibfnamefont{C.-s.} \bibnamefont{Zha}},
  \bibinfo{author}{\bibfnamefont{R.~E.} \bibnamefont{Cohen}},
  \bibinfo{author}{\bibfnamefont{H.-K.} \bibnamefont{Mao}}, \bibnamefont{and}
  \bibinfo{author}{\bibfnamefont{R.~J.} \bibnamefont{Hemley}},
  \bibinfo{journal}{Proceedings of the National Academy of Sciences}
  \textbf{\bibinfo{volume}{111}}, \bibinfo{pages}{4792} (\bibinfo{year}{2014}).

\bibitem[{\citenamefont{Hemley et~al.}(1990{\natexlab{c}})\citenamefont{Hemley,
  Mao, Finger, Jephcoat, Hazen, and Zha}}]{hemley1990equation}
\bibinfo{author}{\bibfnamefont{R.}~\bibnamefont{Hemley}},
  \bibinfo{author}{\bibfnamefont{H.}~\bibnamefont{Mao}},
  \bibinfo{author}{\bibfnamefont{L.}~\bibnamefont{Finger}},
  \bibinfo{author}{\bibfnamefont{A.}~\bibnamefont{Jephcoat}},
  \bibinfo{author}{\bibfnamefont{R.}~\bibnamefont{Hazen}}, \bibnamefont{and}
  \bibinfo{author}{\bibfnamefont{C.}~\bibnamefont{Zha}},
  \bibinfo{journal}{Physical Review B} \textbf{\bibinfo{volume}{42}},
  \bibinfo{pages}{6458} (\bibinfo{year}{1990}{\natexlab{c}}).

\bibitem[{\citenamefont{Akahama et~al.}(2010)\citenamefont{Akahama, Nishimura,
  Kawamura, Hirao, Ohishi, and Takemura}}]{akahama2010evidence}
\bibinfo{author}{\bibfnamefont{Y.}~\bibnamefont{Akahama}},
  \bibinfo{author}{\bibfnamefont{M.}~\bibnamefont{Nishimura}},
  \bibinfo{author}{\bibfnamefont{H.}~\bibnamefont{Kawamura}},
  \bibinfo{author}{\bibfnamefont{N.}~\bibnamefont{Hirao}},
  \bibinfo{author}{\bibfnamefont{Y.}~\bibnamefont{Ohishi}}, \bibnamefont{and}
  \bibinfo{author}{\bibfnamefont{K.}~\bibnamefont{Takemura}},
  \bibinfo{journal}{Phys. Rev. B} \textbf{\bibinfo{volume}{82}},
  \bibinfo{pages}{060101} (\bibinfo{year}{2010}).

\bibitem[{\citenamefont{Bloembergen et~al.}(1967)\citenamefont{Bloembergen,
  Bret, Lallemand, Pino, and Simova}}]{bloembergen1967controlled}
\bibinfo{author}{\bibfnamefont{N.}~\bibnamefont{Bloembergen}},
  \bibinfo{author}{\bibfnamefont{G.}~\bibnamefont{Bret}},
  \bibinfo{author}{\bibfnamefont{P.}~\bibnamefont{Lallemand}},
  \bibinfo{author}{\bibfnamefont{A.}~\bibnamefont{Pino}}, \bibnamefont{and}
  \bibinfo{author}{\bibfnamefont{P.}~\bibnamefont{Simova}},
  \bibinfo{journal}{IEEE Journal of Quantum Electronics}
  \textbf{\bibinfo{volume}{3}}, \bibinfo{pages}{197} (\bibinfo{year}{1967}).

\end{thebibliography}


\end{document}


\preprint{APS/123-QED}

\vspace*{-1in}
\title{Supplementary Material \--- The Raman signal from a hindered hydrogen rotor}
\author{Peter I. C. Cooke$^1$, Ioan B Magdău$^{1,2}$, Miriam Pena-Alvarez$^1$,  Veronika Afonina$^1$, Philip Dalladay-Simpson$^3$, Xiao-Di Liu$^4$, Ross Howie$^3$, Eugene Gregoryanz$^{1,3,4}$,and Graeme J Ackland$^1$}
\affiliation{CSEC, SUPA, School of Physics and Astronomy, The University of Edinburgh, Edinburgh EH9 3JZ, United Kingdom}
\affiliation{Department of Chemistry and Chemical Engineering, Caltech,  Pasadena, California}
\affiliation{Center for High Pressure Science and Technology Advanced Research, Shanghai,China}
\affiliation{Key Laboratory of Materials Physics, Institute of Solid State Physics,  CAS, Hefei, China}

\email{i.b.magdau@sms.ed.ac.uk, gjackland@ed.ac.uk}
\pacs{61.50.Ah, 61.66.Bi, 62.50.-p, 67.80.ff}

\maketitle

\section{Indirect Calculation of Raman Spectra}

In section II.D of the main text a method for obtaining the Raman signal from the response of the system to a sudden excitation was presented. A more standard approach is to generate the Raman signal with an 'indirect method'. In this method the Raman signal is considered to be made up of a series of peaks corresponding to each allowed transition in the system. Continuing from eq 21 the polarisability matrix is given by:

\begin{equation}
\Pi_{ij,lml'm'} = \langle lm \rvert \mathbf{R}^T(\theta,\phi)\cdot \boldsymbol{\alpha}\cdot\mathbf{R}(\theta,\phi) \lvert l'm' \rangle
\end{equation}
where
\begin{equation}
\boldsymbol{\alpha} = 
\left(
\begin{matrix}
1 & 0 & 0 \\
0 & 1 & 0 \\
0 & 0 & \alpha \\
\end{matrix}
\right)
\end{equation}

The peak height corresponding to each individual transition is given by:

\begin{equation}
I_{nn'} = \Pi_{nn'}\rho_{nn'} 
\end{equation}

where

\begin{equation}
\Pi_{nn'} = \sum_{i,j=XY}|\Pi_{ij,nn'}|^2
\end{equation}

for crystallites with c axis parallel to Z or:

\begin{equation}
\Pi_{nn'} = \sum_{i,j=XY}<|\Pi_{ij,nn'}|^2>
\end{equation}

where angled brackets indicate a rotational averaging over all $\theta$ and $\phi$ orientations.

with forbidden transitions evaluating to zero. The corresponding Raman shift for each transition is given by:

\begin{equation}
(\nu_{nn'} - \nu_{0}) = E_{n'} - E_{n}
\end{equation}

To build the overall shape of the spectra the transitions were binned in discrete frequency increments, $\Delta\nu$, and a Lorentzian profile of the form:

\begin{equation}
\phi(\nu) = \frac{I_{i}\frac{\Gamma}{{4\pi^2}}}{(\nu - \nu_{0})^2 + (\frac{\Gamma}{4\pi})^2}
\end{equation}

was overlaid on each bin, where $\Gamma$ is the same parameter used in eq. 36 and $I_i$ is the sum of all $I_{nn'}$ in frequency bin $i$.

\section{\label{sec:level1}EQQ Mean Field Potential}

A quadrupole-quadrupole potential was generated by assuming that there is only pairwise correlation in the orientations of the H$_{2}$ molecules. To generate the potential energy surface the full four-dimensional potential surface was initially calculated for the central rotor in an HCP unit cell with each of it's neighbours in turn using the following expression:

\begin{equation}
V^{EQQ}_{ij}(\theta_1, \phi_1, \theta_2, \phi_2)  = \frac{\Theta^2}{4\pi\epsilon{_0}R_{ij}^5}\Gamma(\theta_1, \phi_1, \theta_2, \phi_2)
\end{equation}

Here $\Theta$ is the quadrupole moment of the $H_2$ or $D_2$ molecule, $R_{ij}$ is the intermolecular separation and $\Gamma$ is the standard angular dependent term given by:

\begin{eqnarray*} \label{eq:eqq-orient}
\Gamma({\bf n}_i, {\bf n}_j, \hat{\vec{R}}) = \frac{3}{4}[35 ({\bf \hat n}_i\cdot\hat{\vec{R}})^2({\bf \hat n}_j\cdot\hat{\vec{R}})^2 -5({\bf\hat n}_i\cdot\hat{\vec{R}})^2  \\
-5({\bf\hat n}_j\cdot\hat{\vec{R}})^2
 - 20({\bf\hat n}_i\cdot\hat{\vec{R}})({\bf\hat n}_j\cdot\hat{\vec{R}})({\bf\hat n}_i\cdot{\bf\hat n}_j)
+ 2({\bf\hat n}_i\cdot{\bf\hat n}_j)^2 + 1]
\end{eqnarray*}

Where ${\bf n}_i$, is the bond orientation vector for molecule $i$, ${\bf n}_j$ is the bond orientation vector for molecule $j$ and $\hat{\vec{R}}$ is the vector from molecule $i$ to molecule $j$.

A Boltzmann weighted average of the orientations of the second rotor was then taken over $\theta_2$, $\phi_2$, giving a two dimensional energy surface:

\begin{equation}
\bar{V}^{EQQ}_{ij}(\theta_1, \phi_1)  = {\frac{\int V^{EQQ}_{ij}\exp{\frac{-V^{EQQ}_{ij}}{k_BT}}}{Z}}\sin\theta_2d\theta_2d\phi_2
\end{equation}
where:

\begin{equation}
Z  = \int{\exp{\frac{-V^{EQQ}_{ij}}{k_bT}}}\sin\theta_2d\theta_2d\phi_2
\end{equation}


Finally the above expression was summed over all pairs to give the molecular field EQQ Potential:

\begin{equation}
V^{M.F.EQQ}(\theta_1, \phi_1)  = \sum_{ij}{\bar{V}^{EQQ}_{ij}(\theta_1, \phi_1)}
\end{equation}\textbf{}

Surprisingly, this EQQ potential has rather little angular dependence. This is because of the absence of three-body correlations and associated frustration.  It implies that in hydrogen EQQ is not strong enough to produce a broken symmetry phase II with a (mean field) single-molecule basis: we note that all candidates for phase II found in DFT have multi-molecular unit cells.

\begin{table}[H]
\centering
\begin{tabular}{|c|c||c|c||c|c||c|c|}
\hline
\multicolumn{2}{|c|}{\bfseries $\mathbf{H_{2}}$, 10 K } & \multicolumn{2}{|c|}{\bfseries $\mathbf{H_{2}}$, 300 K } &
\multicolumn{2}{|c|}{\bfseries $\mathbf{D_{2}}$, 10 K } & \multicolumn{2}{|c|}{\bfseries $\mathbf{D_{2}}$, 300 K }\\
\hline
P & r & P & r & P & r & P & r\\
(GPa) & (\AA) & (GPa) & (\AA) &(GPa) & (\AA) & (GPa) & (\AA)\\
\hline
6.5 &  0.742 & 7.1 & 0.75 & 0.6 &  0.741 & 7.9 & 0.741 \\
15.5  &  0.735 & 13.4 & 0.742 & 2.8 &  0.734 & 12.7 & 0.739\\
31.1 &  0.734 & 19.8 & 0.741 & 7.8 &  0.727 & 22.2 & 0.726\\
45.5 &  0.721 & 30.9 & 0.737 & 15.9 &  0.720 & 32.0 & 0.723\\
54.0 &  0.720 & 48.3 & 0.730 & 21.5 &  0.720 & 49.0 & 0.721\\
\hline
\end{tabular}
\caption{Fitted bond length scaling in hydrogen at various pressures. Fitted values were limited to within 5\% of the gas phase value\cite{Stoicheff1957}.
}
\label{Supp:bondlength}
\end{table}

\begin{table}[H]
\centering
\begin{tabular}{|c|c||c|c||c|c||c|c|}
\hline
\multicolumn{2}{|c|}{\bfseries $\mathbf{H_{2}}$, 10 K } & \multicolumn{2}{|c|}{\bfseries $\mathbf{H_{2}}$, 300 K } &
\multicolumn{2}{|c|}{\bfseries $\mathbf{D_{2}}$, 10 K } & \multicolumn{2}{|c|}{\bfseries $\mathbf{D_{2}}$, 300 K }\\
\hline
P & $\Gamma$ & P & $\Gamma$ & P & $\Gamma$ & P & $\Gamma$ \\
(GPa) & ($cm^{-1}$) & (GPa) & ($cm^{-1}$) & (GPa) & ($cm^{-1}$) & (GPa) & ($cm^{-1}$)\\
\hline
6.5 &  137.9 & 7.1 & 190.7 & 0.5 &  53.1 & 7.9 & 144.6 \\
15.5  &  232.3 & 13.4 & 297.8 & 2.8 &  70.3 & 12.7 & 202.6 \\
31.1 &  396.8 & 19.8 & 406.6 & 7.8 &  107.4 & 22.2 & 316.5 \\
45.5 &  225.0 & 30.9 & 595.3 & 15.9 &  167.5 & 32.0 & 434.2\\
54.0 &  225.0 & 48.3 & 700.00 & 21.5 &  209.1 & 49.0 & 495.0\\
\hline
\end{tabular}
\caption{Fitted values for parameter $\Gamma$ (see eq 36 in main text) for hydrogen and deuterium at 10 K and 300 K.}
\label{Supp:GammaTable}
\end{table}

\begin{table}[H]
\centering
\begin{tabular}{|c|c||c|c|}
\hline
\multicolumn{2}{|c||}{\bfseries $\mathbf{H_{2}}$, 10 K } & \multicolumn{2}{|c|}{\bfseries $\mathbf{D_{2}}$, 10 K } \\
\hline
P (GPa) & o:p & P (GPa) & o:p \\
\hline
5.4 & 75:25 & 0.6 & 70:30 \\
6.5 & 75:25 & 2.8 & 70:30 \\
10.1 & 75:25 & 7.8 & 70:30   \\
15.5 & 75:25 & 15.9 & 80:20 \\
19.3 & 70:30 & 21.5 & 80:20 \\
24.5 & 65:35 & 31.8 & - \\
31.1 & 50:50 & 40.1 & - \\
35 & 32:68 & 48.1 & - \\
45.5 & 25:75 & 54.5 & - \\
54 & 20:80 & 58.0 & - \\
\hline
\end{tabular}
\caption{$ortho$-$para$ ratios for hydrogen and deuterium fitted to experimental intensities at 10 K. At 300 K the ratio was calculated directly from the temperature (i.e. $T'=T$).
}
\label{Supp:opratio}
\end{table}

\begin{figure*}[h!]
\includegraphics[width=0.5\linewidth]{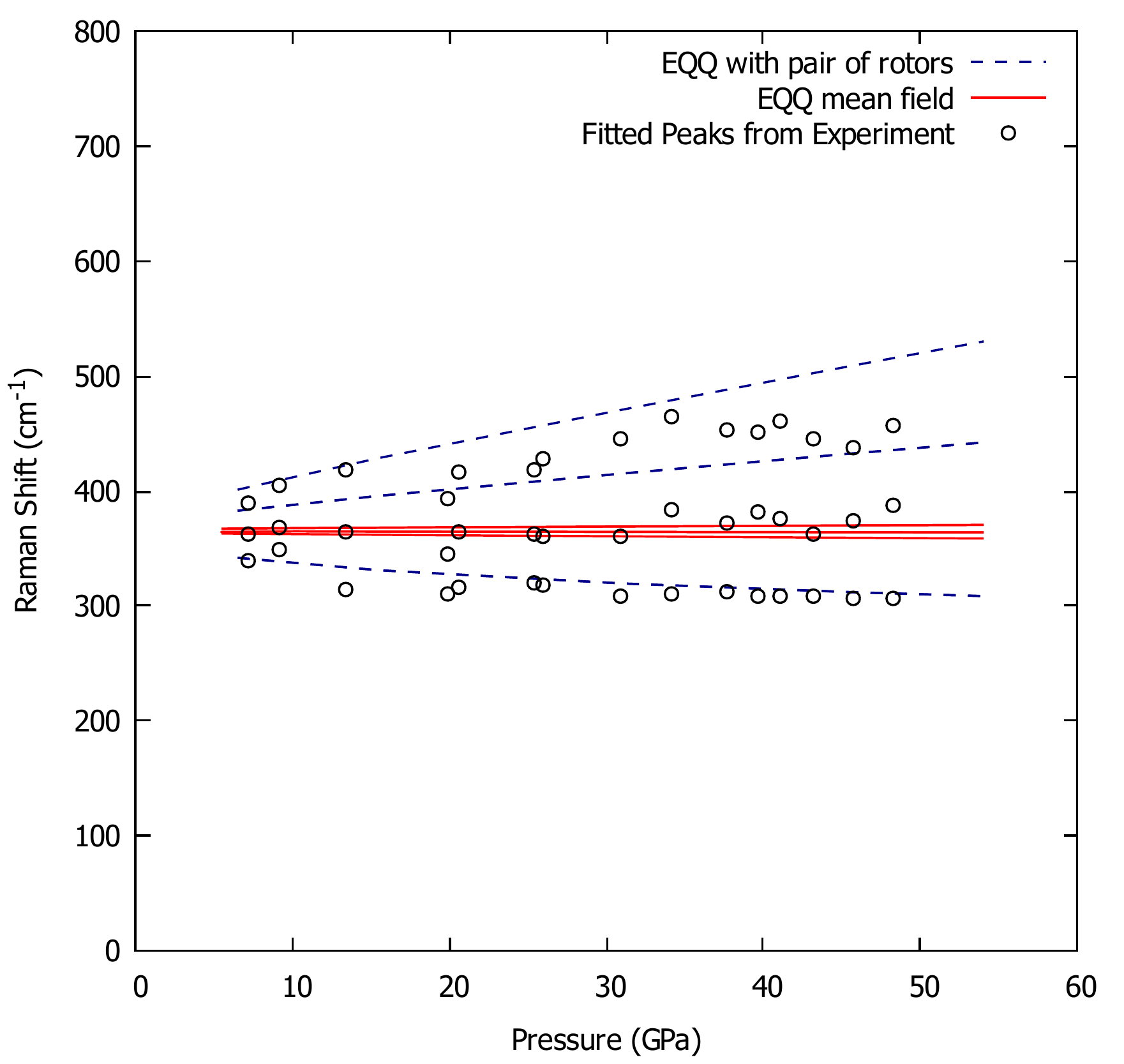}
\caption{Frequency shifts for the $S_0(0)$ triplet are shown for the mean field EQQ model (red), the EQQ interaction with a second rotor with a fixed orientation of $\theta=90, \phi=0$ (blue dashed) and the least squares fitted experimental peaks (black circles). The mean field only generates small differences in frequency between the $S_0(0)$ contributions, in stark contrast to the fitted shifts from experiment. The quadrupolar interaction with a single rotor at fixed orientation creates a larger difference in frequency between $S_0(0)$ contributions but still shows stark disagreement with experiment at 50 GPa and is not a representative model of the HCP crystal in phase I. This demonstrates that while there is sufficient energy in the quadrupole interaction to create the necessary splitting a 'parameter-free' mean field model alone cannot explain the splitting in the $S_0(0)$ triplet, as shown in previous theoretical work on the solid phases of hydrogen.\cite{Freiman2005}}
\label{Supp:EvsP_EQQComparison}
\end{figure*}

\begin{figure*}[p!]
\includegraphics[width=0.7\linewidth]{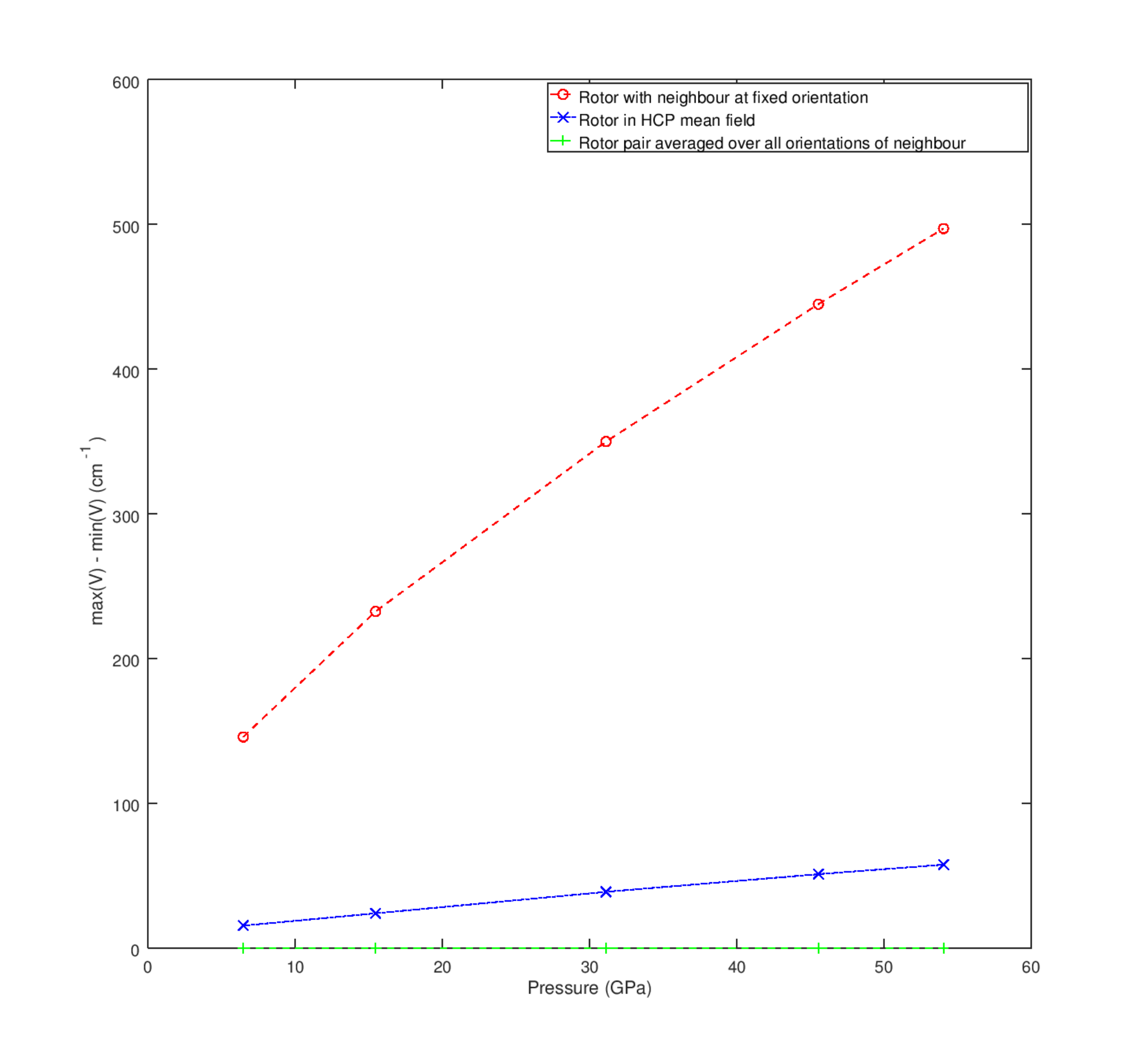}
\caption{Difference in energy between the minima and maxima of the potential energy surface for a quadrupole in 3 different cases are shown. A quadrupole with a neighbour of fixed orientation $(\theta = \frac{\pi}{2}, \phi = 0)$ is shown in red. A quadrupole in an HCP mean field is shown in blue and a quadrupole with a completely uncorrelated neighbour (averaged over all orientations) is shown in green. The mean field shows a reduction of $\sim 95\%$ compared to the a single neighbour at fixed orientation, demonstrating that a parameter free mean field approach is not suitable and hence 3 free parameters were introduced to fit a potential consisting of long range electrostatic interactions and short range steric repulsion.} 
\label{Supp:DeltaV}
\end{figure*}

\begin{figure*}[h!]
\includegraphics[width=0.45\linewidth]{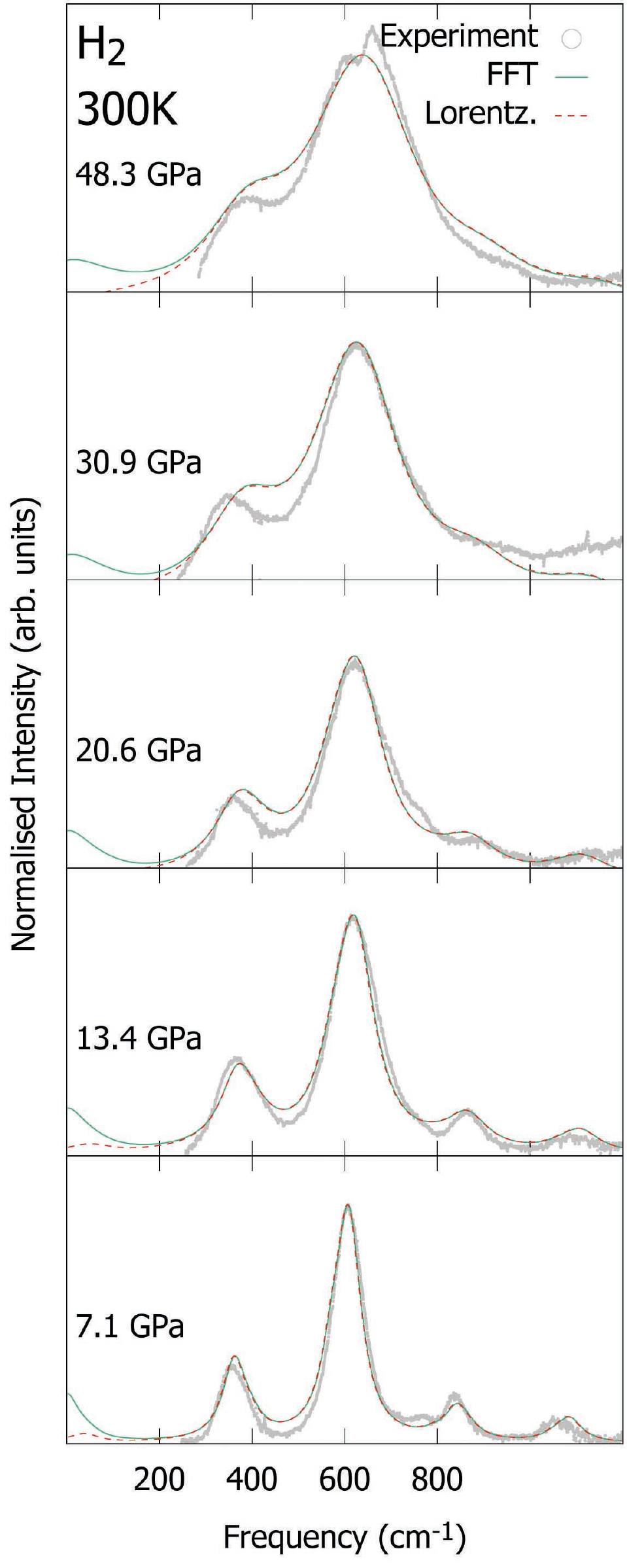}
\caption{Calculated Raman spectra using direct(FFT) and indirect (Lorentzians) methods for hydrogen at 300 K. Excellent agreement between both methods can be seen. The divergence seen at low frequencies is due to a different method of filtering out Rayleigh transitions in each method. Perfect agreement could be achieved with a simple adaptation to the Rayleigh filtering method in the FFT approach in future work.}
\label{Supp:FFTvLorentzH2_300K}
\end{figure*}

\begin{figure*}[h!]
\includegraphics[width=0.45\linewidth]{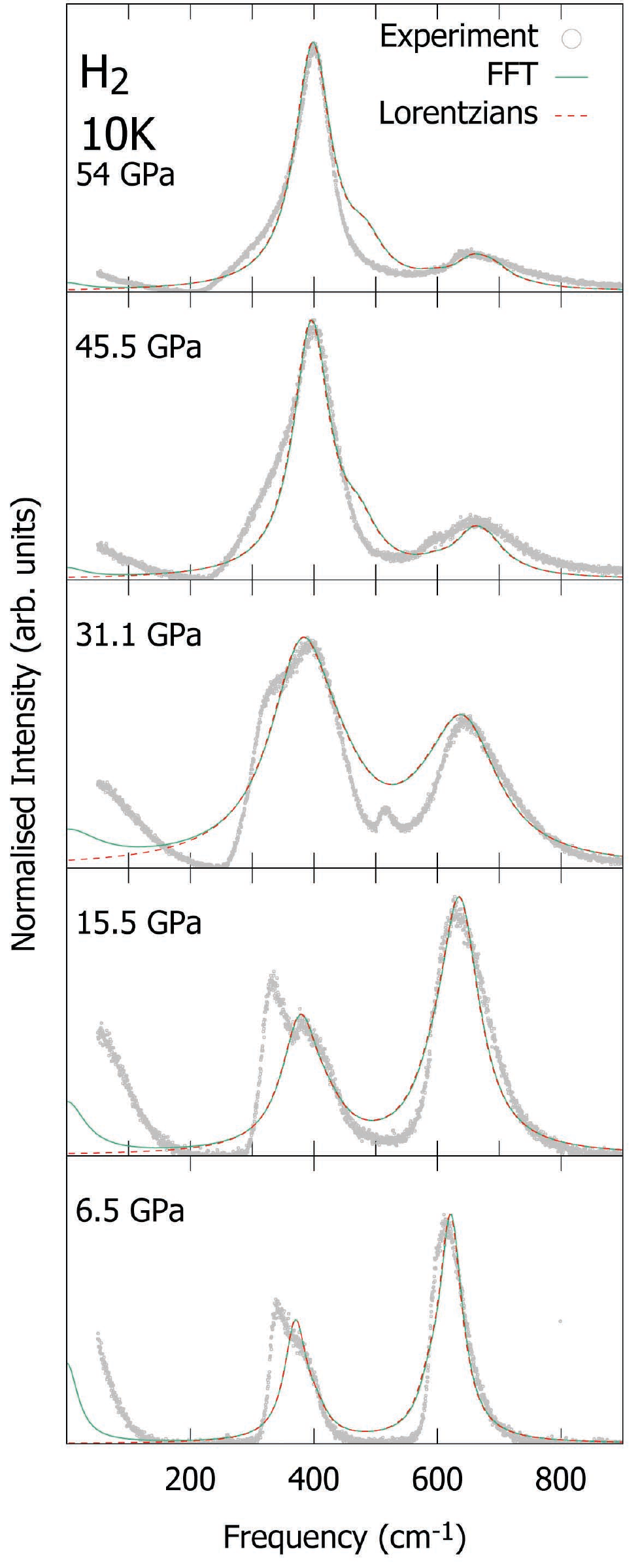}
\caption{Calculated Raman spectra using direct(FFT) and indirect (Lorentzians) methods for hydrogen 10 K.}
\label{Supp:FFTvLorentzH2_10K}
\end{figure*}

\begin{figure*}[h!]
\includegraphics[width=0.45\linewidth]{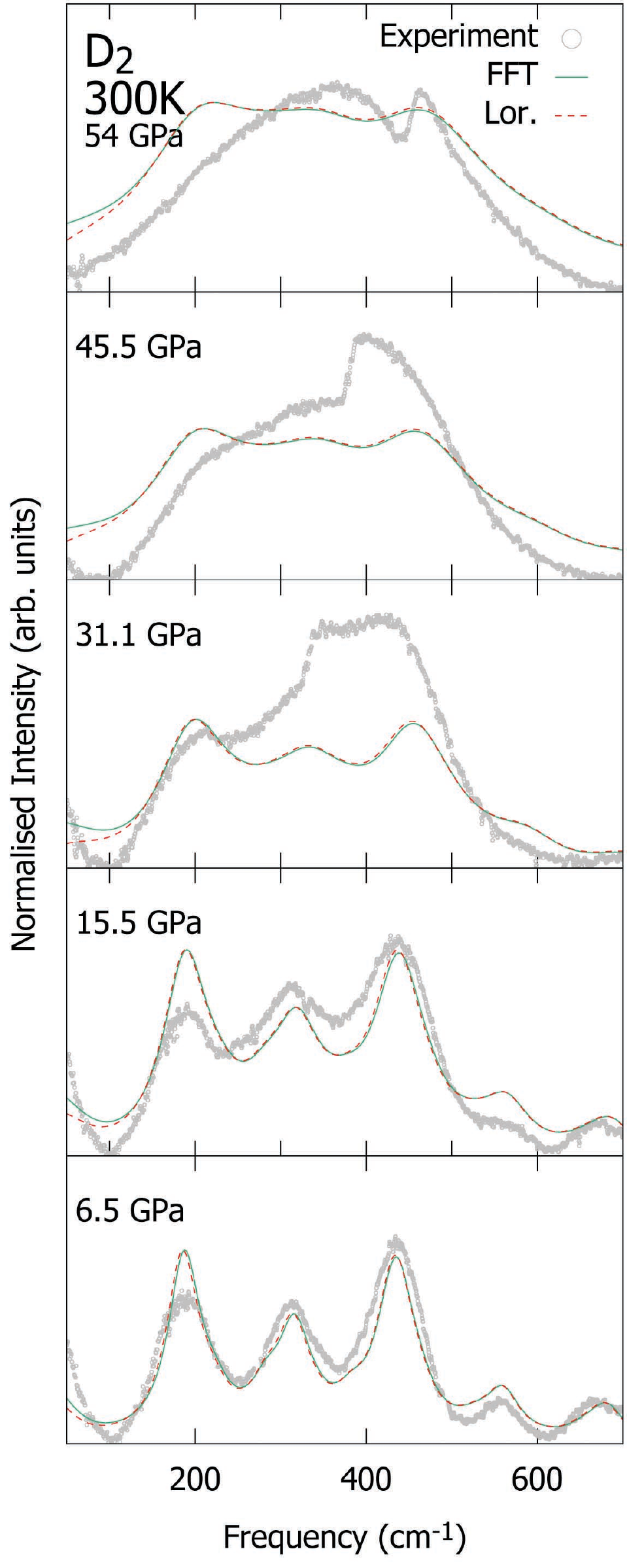}
\caption{Calculated Raman spectra using direct(FFT) and indirect (Lorentzians) methods for deuterium at 300 K.}
\label{Supp:FFTvLorentzD2_300K}
\end{figure*}

\begin{figure*}[h!]
\includegraphics[width=0.45\linewidth]{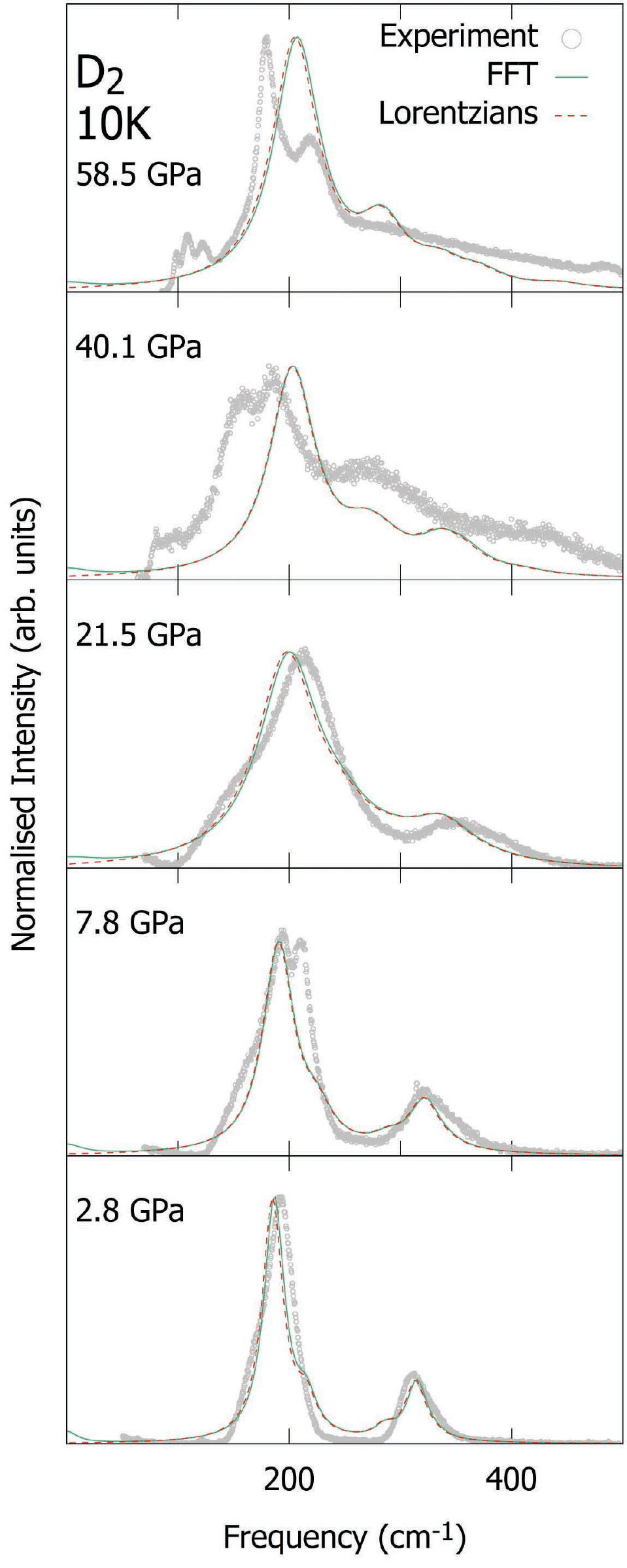}
\caption{Calculated Raman spectra using direct(FFT) and indirect (Lorentzians) methods for deuterium at 10 K.}
\label{Supp:FFTvLorentzD2_10K}
\end{figure*}

\begin{figure*}[h!]
\includegraphics[width=0.4\linewidth]{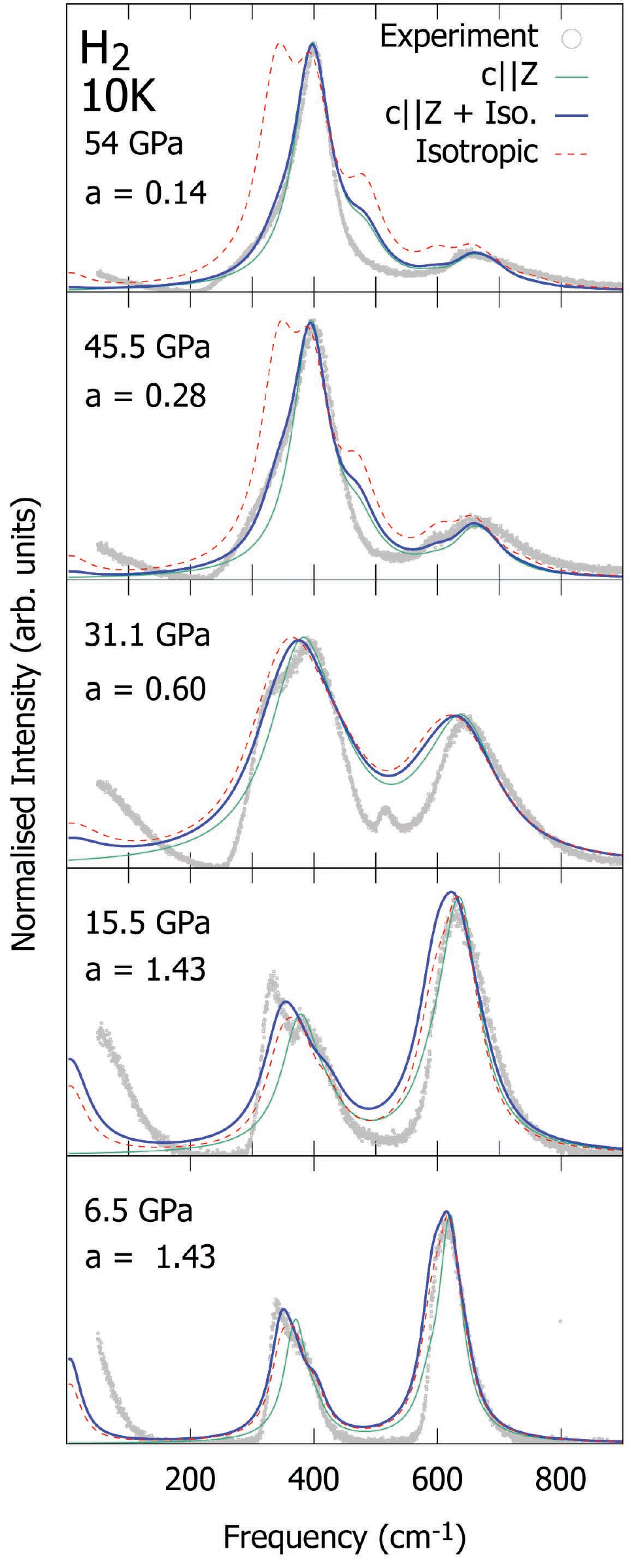}
\caption{We also include a signal corresponding to a combination of both geometries (blue). In this case the overall signal, $f(\omega)$, is generated using a weighting parameter 'a' where $f(\omega) = af_{isotropic}(\omega) + (1-a)f_{c||Z}(\omega)$. The signal is much more accurately described by a linear combination of the two geometries at higher spin temperatures i.e. when a higher proportion of $ortho$-hydrogen molecules are present in the sample. The close agreement with experiment achieved here is consistent with the observation that the intensity of the $|m_j|=1$ peak in the $S_0(0)$ triplet is very well correlated to the abundance of $ortho$-hydrogen as seen in previous experimental studies \cite{Eggert1999}.}.
\label{Supp:AparamH210K}
\end{figure*}

\begin{table}[H]
\centering
\begin{tabular}{|c|c|c|c|}
\hline
 & $|m_j|=1$ & $|m_j|=2$ & $|m_j|=0$ \\
\hline
$S_{c||Z}$ & 0 & 12 & 1 \\
$S_{iso.}$ & 2 & 2 & 1 \\
$S_{a}$ & 2a & 12-10a & 1 \\
\hline
\end{tabular}
\caption{Fitted $ortho$-$para$ ratios for hydrogen and deuterium at 10 K. At 300 K the ratio was calculated directly from the temperature.
}
\label{TripletIntensities}
\end{table}

\begin{figure*}[h!]
\includegraphics[width=0.4\linewidth]{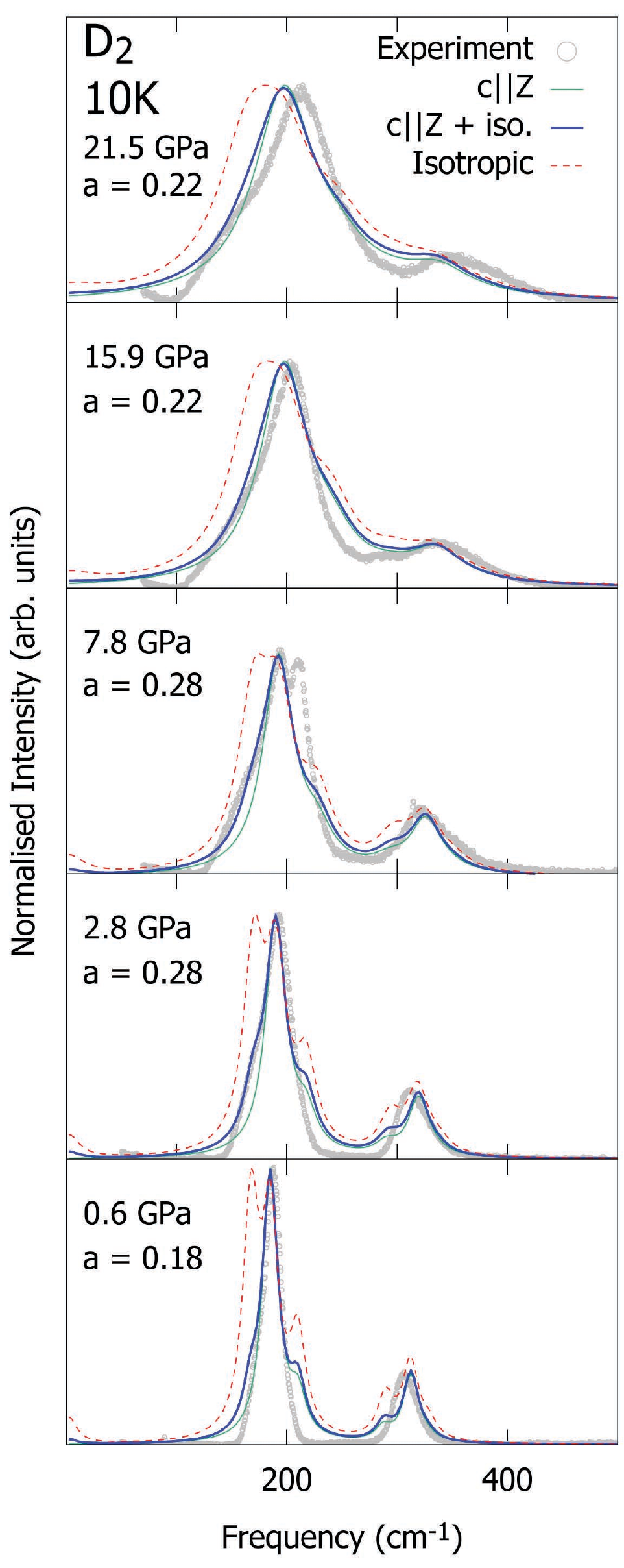}
\caption{Deuterium spectra at 10 K, fitted with the procedure described in fig \ref{Supp:AparamH210K}. The 'a' parameter shows a much smaller variation with increasing time/pressure with respect to hydrogen. This is consistent with the relatively small change in $ortho$-deuterium fraction compared to hydrogen. (See Table \ref{Supp:opratio})}
\label{Supp:AparamD210K}
\end{figure*}

\begin{figure*}[h!]
\includegraphics[width=0.5\linewidth]{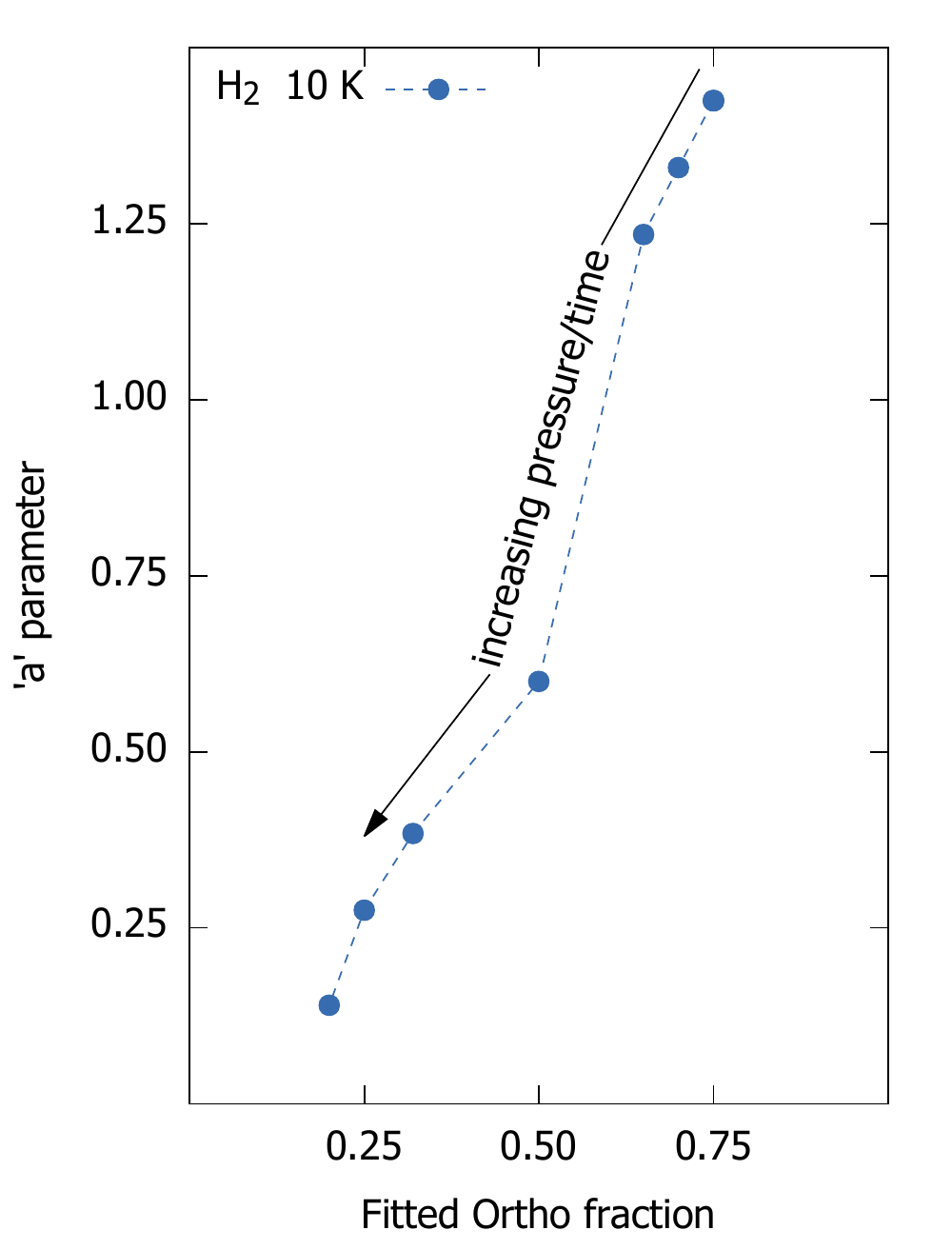}
\caption{
Variation of the a parameter and the $ortho$-$para$ ratio used to fit the experimental data at 10 K.  The $ortho$-$para$ ratio is used to fit the relative peak heights of $S_0(0)$ and $S_0(1)$, corresponding to the occupation of the J=0 and J=1 energy levels.  The "a" parameter is a preferred-orientation correction required to fit the $S_0(0)$ peak shape.  The strong correlation between the two shows that the "preferred orientation" represents the fact that the M$_J$=1 component is only visible due to resonant scattering via the $ortho$-peaks.    In fact there is always strong alignment with $c||Z$ ,  which means that once the $ortho$-H$_2$ has converted, the M$_J$=2 component of $S_0(0)$ is dominant.
The $ortho$ fraction of 0.75 corresponds to room temperature, and drops to the 10 K equilibrium value as the pressurization proceeds. The calculation suggests that the changing $S_0(0)$ peak shape is mainly due to the $ortho$-$para$ ratio rather than changes in crystal orientations within the cell, or the pressure.
}
\label{Supp:SpinTempvsExp}
\end{figure*}

\clearpage